\documentclass[10pt,aps,pra,showpacs,showkeys,twocolumn]{revtex4-1}
\usepackage{amsmath, amssymb}
\usepackage{amsthm}
\usepackage{graphicx}
\usepackage{cases}
\usepackage{bbold}
\usepackage{xcolor}
\usepackage{booktabs}
\usepackage{dcolumn}
\usepackage{textcomp}
\usepackage{multirow}
\usepackage[customcolors]{hf-tikz}

\DeclareMathOperator{\tr}{Tr}

\DeclareMathOperator{\erf}{erf}

\renewcommand{\vec}[1]{\mathbf{#1}}

\tikzset{style green/.style={
    set fill color=green!50!lime!60,
    set border color=white,
  },
  style cyan/.style={
    set fill color=cyan!90!blue!60,
    set border color=white,
  },
  style orange/.style={
    set fill color=orange!80!red!60,
    set border color=white,
  },
  hor/.style={
    above left offset={-0.15,0.31},
    below right offset={0.15,-0.125},
    #1
  },
  ver/.style={
    above left offset={-0.1,0.3},
    below right offset={0.15,-0.15},
    #1
  }
}

\begin{document}

\title{Generalized conditions for genuine multipartite continuous-variable 
entanglement}
\author{E. Shchukin}
\email{evgeny.shchukin@gmail.com}
\author{P. van Loock}
\email{loock@uni-mainz.de}
\affiliation{Johannes-Gutenberg University of Mainz, Institute of Physics, 
Staudingerweg 7, 55128
Mainz}

\begin{abstract}
We derive a hierarchy of continuous-variable multipartite entanglement 
conditions in terms of second-order moments of position and momentum operators 
that generalizes existing criteria. Each condition corresponds to a convex 
optimization problem which, given the covariance matrix of the state, can be 
numerically solved in a straightforward way. The conditions are independent of 
partial transposition and thus are also able to detect bound entangled states. 
Our approach can be used to obtain an analytical condition for genuine 
multipartite entanglement. We demonstrate that even a special case of our 
conditions can detect entanglement very efficiently. Using multi-objective 
optimization it is also possible to numerically verify genuine entanglement of 
some experimentally realizable states. 
\end{abstract}



\maketitle

\section{Introduction} 

\begin{figure}
\includegraphics[scale=0.4]{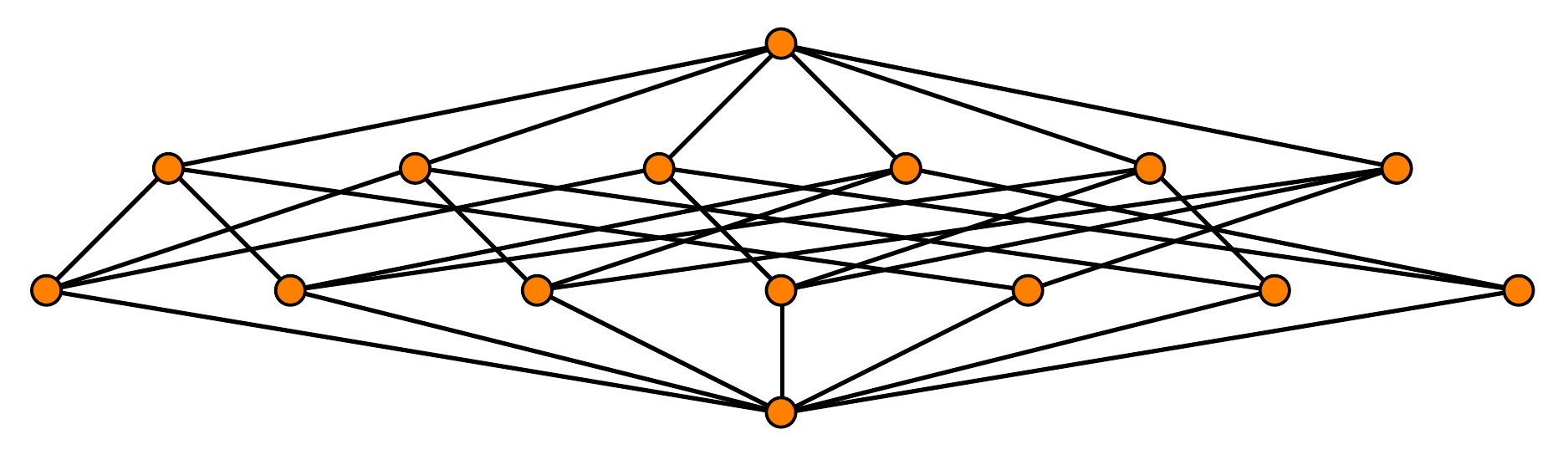}
\caption{(Color online). The partition lattice of four 
parties. Nodes are partitions and edges correspond to the 
"finer than" relation.}\label{fig:partitions}
\end{figure}

In the multipartite case there are many different 
notions of entanglement, ranging from the most specific to the most general 
ones. A specific kind of entanglement means that we precisely specify the 
groups 
of parties that are separable from each other, i.e., we specify a partition of 
the set of indices $\{1, \ldots, n\}$. A partition $\mathcal{I} = \{I_1, 
\ldots, I_k\}$ is a disjoint set of nonempty subsets of the indices whose union 
is equal to $\{1, \ldots, n\}$. For example, in the case $n=4$ there are 
partitions $\{2\} \cup \{1, 3, 4\}$, $\{1\} \cup \{2\} \cup \{3, 4\}$ and many 
others. We will use a more compact notation for partitions and write them as 
$2|134$ and $1|2|34$. A partition $\mathcal{I}'$ is finer than a partition 
$\mathcal{I}$ (or the partition $\mathcal{I}$ is coarser than $\mathcal{I}'$) 
if 
for any $I'_i$ there is an $I_j$ such that $I'_i \subseteq I_j$. For example, 
the partition $1|2|34$ is finer than the partition $12|34$. There are two 
extreme partitions, the trivial partition $1 2 \ldots n$ and the partition 
$1|2|\ldots|n$. The former corresponds to the case where no information about 
separability properties is known and the latter corresponds to the notion of 
full separability, where all parties are separable from each other. The set of 
partitions with the finer-than relation is referred to as partition lattice and 
can be visualized as the graph shown in Fig.~\ref{fig:partitions} for $n=4$. At 
the bottom is the trivial partition $1 \ldots n$, then on the second line are 
all $2^{n-1}-1$ partitions into two parts, next are all partitions into three 
parts, and so on. At the top is the partition $1|2|\ldots|n$. Partitions with 
$k=2$ play a special role and are called bipartitions. 

From these specific kinds of separability one can construct more general types 
by considering mixtures of specific kinds according to some criteria. For 
example, a multipartite state is called $k$-separable, $k \geq 2$, if it is a 
mixture of states each of which has $k$ separable groups of modes (the 
corresponding partitions are on the same line in the lattice diagram). In the 
case of $n=4$, a $3$-separable state is a state which is a mixture of 
$1|2|34$-, $1|3|24$-, $1|4|23$-, $2|3|14$-, $13|2|4$- and $12|3|4$-separable 
states. The notion of $n$-separability of an $n$-partite state is the same as 
the notion of full separability. A $2$-separable state is referred to as 
biseparable and the notion of biseparability is the most general of all --- a 
state which is separable in any sense discussed above is automatically 
biseparable. On the other hand, full separability is the most specific kind of 
separability --- a fully separable state is separable in any other sense. 

In general, a biseparable state is a mixture of $2^{n-1}-1$ states 
corresponding to all nontrivial bipartitions of $n$ parties. There are states 
that cannot be represented as such a mixture. These states, the states that are 
not biseparable, are called genuine miltipartite entangled. The problem of 
recognizing which states are genuine entangled is very important for different 
applications. It is a highly nontrivial task to develop practical conditions 
for detecting genuine multipartite entanglement. In the following a single 
party always corresponds to a single electromagnetic mode as known from 
continuous-variable (CV) quantum information theory \cite{RevModPhys.77.513, 
RevModPhys.84.621}. Many such conditions for $n$-partite CV systems deal with 
lower bounds for the second-order quantity 
\begin{equation}
    \mathcal{G} = \langle \hat{\vec{r}}^{\mathrm{T}} M \hat{\vec{r}}\rangle 
    = \sum^n_{i,j=1} M_{ij} \langle\{\hat{r}_i, \hat{r}_j\}\rangle = \tr(M 
\gamma),
\end{equation}
where $\hat{\vec{r}} = (\hat{\vec{x}}, \hat{\vec{p}})$ is the $2n$-vector of 
position and momentum operators and $M = (M_{ij})^{2n}_{i,j=1}$ is a real, 
symmetric, positive definite $2n \times 2n$ matrix and $\gamma = 
(\langle\{\hat{r}_i, \hat{r}_j\}\rangle)^{2n}_{i,j=1}$ is the covariance matrix 
of the state, $\langle\{\hat{A}, \hat{B}\}\rangle = 
(1/2)\langle\hat{A}\hat{B}+\hat{B}\hat{A}\rangle$. In fact, a special case of 
matrix $M$ 
with zero off-diagonal blocks is usually used in practice. Denoting $X = 
(M_{ij})^n_{i,j=1}$ and $P = (M_{ij})^{2n}_{i,j=n+1}$, which are symmetric, 
positive semidefinite $n \times n$ matrices, the quantity $\mathcal{G}$ in this 
case reads as 
\begin{equation}\label{eq:GXP}
    \mathcal{G} = \langle \hat{\vec{x}}^{\mathrm{T}} X \hat{\vec{x}} + 
\hat{\vec{p}}^{\mathrm{T}} P \hat{\vec{p}}\rangle 
    = \tr(X \gamma_{xx}) + \tr(P \gamma_{pp}),
\end{equation}
where $\gamma_{xx} = (\langle\hat{x}_i\hat{x}_j\rangle)^n_{i,j=1}$ and 
$\gamma_{pp} = (\langle\hat{p}_i\hat{p}_j\rangle)^n_{i,j=1}$ are the diagonal 
blocks of the covariance matrix of the state. For example, the classical result 
of Ref.~\cite{PhysRevLett.84.2722} uses rank-one matrices
$X = \left(
\begin{smallmatrix}
    a^2 & 1 \\
    1 & a^{-2}
\end{smallmatrix}
\right) = \vec{h}\vec{h}^{\mathrm{T}}$
and 
$P = \left(
\begin{smallmatrix}
    a^2 & -1 \\
    -1 & a^{-2}
\end{smallmatrix}
\right) = \vec{g}\vec{g}^{\mathrm{T}}$, where $\vec{h}^{\mathrm{T}} = 
(a,a^{-1})$ and $\vec{g}^\mathrm{T} = (a,-a^{-1})$. The 
works~\cite{PhysRevA.67.052315, PhysRevA.90.062337} use general rank-one 
matrices $X = \vec{h}\vec{h}^{\mathrm{T}}$ and $P = 
\vec{g}\vec{g}^{\mathrm{T}}$, where $\vec{h}$ and $\vec{g}$ are real 
$n$-vectors. In Ref.~\cite{PhysRevA.90.012334} the quantity 
$\mathcal{G}$ with $3 \times 3$ matrices was used. 
Refs.~\cite{PhysRevLett.111.110503, PhysRevLett.114.050501} deal with 
the general quantity $\mathcal{G}$. Among the works that also use second 
order moments we mention \cite{NewJPhys.8.51, NatPhys.9.19, 
PhysRevA.90.052321}. As based on second-order moments only, all these criteria 
are sufficient entanglement witnesses for all CV states, but necessary and 
sufficient only for Gaussian states.

In a realistic setting errors of measurements are unavoidable, so to be 
completely rigorous we need to incorporate the possibility of errors into our 
scheme. We do this and formulate a hierarchy of entanglement conditions as 
convex optimization problems for $\mathcal{G}$ in terms of the 
covariance matrix $\gamma$ and the information about errors of the 
measurements. For discrete variables a convex optimization approach has been 
developed in Ref.~\cite{PhysRevA.70.062317}.

The paper is organized as follows. In section \ref{sec:qb} we obtain the 
minimal value of $\mathcal{G}$ defined by Eq.~\eqref{eq:GXP} over all quantum 
states. In Section \ref{sec:sc} we show how to improve the lower bound obtained 
in the preceding section for separable states and construct a hierarchy of 
separability conditions in the form of convex optimization problems. In Section 
\ref{sec:r1} we apply our construction to rank one matrices and demonstrate that 
it coincides with some previously known results, so that those results are just 
a special case of our more general approach. In Section \ref{sec:ppt} we give 
an example of an entangled state with positive partial transposition that can 
be detected by our conditions. Section \ref{sec:ge} is devoted to an analytical 
condition for genuine multipartite entangled that can be obtained from our 
hierarchy of conditions. in Section \ref{sec:co} we further extend our approach 
by taking measurement errors into consideration and demonstrating that our 
method works in realistic settings as well. Then we give a conclusion and 
provide in appendices all technical details missing in the main part of our 
work.

\section{Quantumness bound}
\label{sec:qb}

First of all, we find the minimum of $\mathcal{G}$ over all quantum states and 
then we show how this bound can be improved for separable states. To find the 
minimal value of $\mathcal{G}$ for a given matrix $M$ we use the Williamson's 
theorem \cite{*[][{ p. 244}] SGQM}. This theorem states that there is a 
symplectic matrix $S$ such that 
$
    S^{\mathrm{T}} M S = 
    \left(
    \begin{smallmatrix} 
        \Lambda & 0 \\ 
        0 & \Lambda 
    \end{smallmatrix}
    \right)
$, 
where $\Lambda$ is a diagonal $n \times n$ matrix. Since each symplectic 
transform is implementable as a unitary transformation \cite{PhysRevA.37.3028}, 
starting with a state $\hat{\varrho}$ we have 
\begin{equation}
    \langle \hat{\vec{r}}^{\mathrm{T}} M \hat{\vec{r}}\rangle = 
    \left\langle \hat{\vec{r}}^{\mathrm{T}} 
    \begin{pmatrix} 
        \Lambda & 0 \\ 
        0 & \Lambda 
    \end{pmatrix} 
    \hat{\vec{r}}\right\rangle' 
\end{equation}
for the appropriately transformed state $\hat{\varrho}'$. The minimum of 
$\mathcal{G}$ 
is achieved for $\hat{\varrho}'$ being the vacuum state, and in this case the 
equality $\mathcal{G} = \tr \Lambda$ takes place. We thus obtain that the 
inequality $\mathcal{G} = \tr(M\gamma) \geq \tr \Lambda$ is valid for all 
quantum states and it is tight.

To compute the minimal value of $\mathcal{G}$ we need to know the diagonal 
elements $\lambda_j$ of $\Lambda$. These numbers are referred to as 
\textit{symplectic spectrum} of $M$ and they can be 
directly obtained from the matrix $M$ according to the fact that 
$\pm i \lambda_j$ are the eigenvalues of $J M$, where 
$
    J = 
    \left(
    \begin{smallmatrix}
        0 & E \\
        -E & 0
    \end{smallmatrix}
    \right),
$
and $E$ is the identity matrix. In our special case of block-diagonal 
$
    M = 
    \left(
    \begin{smallmatrix} 
        X & 0 \\ 
        0 & P 
    \end{smallmatrix}
    \right)
$
we can get these numbers directly in terms of $X$ and $P$. In fact, we have the 
equality
\begin{equation}
    J M = 
    \begin{pmatrix}
        0 & E \\
        -E & 0
    \end{pmatrix}
    \begin{pmatrix}
        X & 0 \\
        0 & P
    \end{pmatrix}
    =
    \begin{pmatrix}
        0 & P \\
        -X & 0
    \end{pmatrix}.
\end{equation}
The characteristic equation of this matrix is 
$
    \chi(\lambda) = 
    \det
    \left(
    \begin{smallmatrix}  
        \lambda E & -P \\ 
        X & \lambda E
    \end{smallmatrix}
    \right) = 0
$.
Since the diagonal blocks commute with the off-diagonal ones, according to 
\cite{*[][{ p. 27, Eq.~(0.8.5.13)}] horn-johnson} this equation can be 
simplified as $\chi(\lambda) = \det(\lambda^2 E + XP) = 0$. Substituting the 
eigenvalues $\pm i \lambda_j$ into this equation we see that the diagonal 
elements $\lambda_j$ satisfy the equation 
\begin{equation}
    \det(XP - \lambda^2 E) = \det(\sqrt{X}P\sqrt{X} - \lambda^2 E) = 0,
\end{equation} 
from which it follows that they are the eigenvalues of the symmetric matrix 
$\sqrt{\sqrt{X}P\sqrt{X}}$. The matrices $X$ and $P$ can be swapped in this 
derivation. We arrive to the following result: The minimal value of 
$\mathcal{G}$ is given by the expression
\begin{equation}\label{eq:minG2}
    \min_{\hat{\varrho}} \mathcal{G} = \tr \sqrt{\sqrt{X}P\sqrt{X}} 
    = \tr \sqrt{\sqrt{P}X\sqrt{P}}.
\end{equation}
To put it in another way, we have the tight inequality 
\begin{equation}\label{eq:minG3}
    \tr(X \gamma_{xx}) + \tr(P \gamma_{pp}) \geq 
    \mathcal{B}(X, P) = \tr\sqrt{\sqrt{X}P\sqrt{X}},
\end{equation}
which is valid for all positive semidefinite matrices $X$ and $P$ and all 
correlation matrices $\gamma_{xx}$ and $\gamma_{pp}$. This inequality 
gives some bound on $\mathcal{G}$. The bound has been 
obtained without any assumptions about separability properties of the state 
in question and thus is valid for all multipartite quantum states. Our task now 
is to improve this bound for separable states. The improvement will depend on 
the separability kind of the state --- the more separable groups of modes the 
state has, the higher is its separability kind in the partition lattice and the 
more can this estimation be improved. In Appendix A we also show 
that from the inequality \eqref{eq:minG3} one can get a special case of 
Araki-Lieb-Thirring trace inequalities \cite{PhysRevLett.35.687, lieb-thirring, 
Lett.Math.Phys.19.167}. Such inequalities play some role in quantum entropy 
theory, see Ref.~\cite{entropy}.

\section{Hierarchy of separability conditions} 
\label{sec:sc}

We first show that pure states with real wave functions are enough to minimize 
$\mathcal{G} = \mathcal{G}(X, P)$. In fact, in terms of wave functions the 
quantity $\mathcal{G}$ reads as
\begin{equation}\label{eq:G}
    \mathcal{G} = \int \Bigl((\vec{x}^{\mathrm{T}} X \vec{x}) 
|\psi(\vec{x})|^2 + (\nabla\psi^*)^{\mathrm{T}} P (\nabla\psi)\Bigr) \, 
d\vec{x}.
\end{equation}
If we take a general wave function of the form $\psi(\vec{x}) = f(\vec{x}) e^{i 
\varphi(\vec{x})}$, where $f(\vec{x}) = |\psi(\vec{x})|$ is a real wave 
function and $\varphi(\vec{x})$ is the phase, we will see that the first term 
in Eq.~\eqref{eq:G} does not depend on $\varphi(\vec{x})$, while the 
other term is equal to 
\begin{equation}
\begin{split}
    \int &((\nabla f)^{\mathrm{T}} P (\nabla f) + 
f^2 (\nabla\varphi)^{\mathrm{T}} P (\nabla\varphi)) \, d\vec{x} \\
    &\geq
    \int (\nabla f)^{\mathrm{T}} P (\nabla f) \, d\vec{x}.
\end{split}
\end{equation}
It follows that for any wave function $\psi(\vec{x})$ there is a real wave 
function $f(\vec{x})$ (the absolute value of the former) such that
$\mathcal{G}_\psi(X, P) \geq \mathcal{G}_f(X, P)$ and thus we can consider only 
pure states with real wave functions.

If a pure state with real wave function is $\{I_1, \ldots, I_k\}$-separable 
then $\gamma_{pp, ij} = \langle\hat{p}_i\hat{p}_j\rangle = 0$ if the indices $i$ 
and $j$ belong to different blocks $I_l$. In fact, separability for pure states 
means that the wave function is factorizable, i.e., $\psi(\vec{x}) = 
f(\vec{x}') g(\vec{x}^{\prime\prime})$, where, without loss of generality, we 
can assume that $\vec{x}' = (x_1, \ldots, x_k)$, $\vec{x}^{\prime\prime} = 
(x_{k+1}, \ldots, x_n)$. For $1 \leq i \leq k$ and $k+1 \leq j \leq n$ we 
have 
\begin{equation}
\begin{split}
    &\langle\hat{p}_i\hat{p}_j\rangle = \int 
    \frac{\partial \psi}{\partial x_i}(\vec{x}) 
    \frac{\partial \psi}{\partial x_j}(\vec{x}) \, d\vec{x} \\
    &= \int f(\vec{x}') \frac{\partial f}{\partial x_i}(\vec{x}') 
    \, d\vec{x}' \int g(\vec{x}^{\prime\prime})  
    \frac{\partial g}{\partial x_j}(\vec{x}^{\prime\prime}) \, 
    d\vec{x}^{\prime\prime} = 0.
\end{split}
\end{equation}
The fact that $\psi(\vec{x})$ is real is important for the validity of the last 
step. The seeming asymmetry in position and momentum operators is superficial 
--- if we worked in momentum representation and dealt with real wave function 
in that representation we would have $\langle\hat{x}_i\hat{x}_j\rangle = 0$.

We have just shown that $\langle\hat{p}_i\hat{p}_j\rangle = 0$ if modes $i$ and 
$j$ are separable, but for the position correlations we can say only that they 
factorize: $\langle\hat{x}_i\hat{x}_j\rangle = \langle\hat{x}_i\rangle 
\langle\hat{x}_j\rangle$. We can get a similar conclusion for position 
moments if we take minimization property into account. If a state 
$|\psi\rangle$ minimizes $\mathcal{G}$ then we can also assume that 
$\langle\hat{x}_i\hat{x}_j\rangle = 0$. In fact, taking the wave function 
$\psi_0(\vec{x}) = \psi(\vec{x} + \vec{x}_0)$, where $\vec{x}_0 = 
\langle\vec{\vec{x}}\rangle$ is the vector of averages computed for the wave 
function $\psi(\vec{x})$, we get a new wave function with 
$\langle\vec{\vec{x}}\rangle_0 = \vec{0}$ and $\mathcal{G}_0 = \mathcal{G} - 
\langle\vec{\vec{x}}\rangle^{\mathrm{T}} X \langle\vec{\vec{x}}\rangle \leq 
\mathcal{G}$. If $X$ is positive definite then we must have 
$\langle\vec{x}\rangle = \vec{0}$ and thus $\langle\hat{x}_i\hat{x}_j\rangle = 
0$ for any separable state $\psi(\vec{x})$ that minimizes $\mathcal{G}$. If $X$ 
is degenerate then we can just find a separable state with 
$\langle\hat{x}_i\hat{x}_j\rangle = 0$ that minimizes $\mathcal{G}$, but there 
can be minimizing states that do not have this property. Moreover, this new wave 
function $\psi_0(\vec{x})$ also has the property that 
$\langle\hat{p}_i\hat{p}_j\rangle = 0$ if $i$ and $j$ are separable. We see that 
among $\{I_1, \ldots, I_k\}$-separable states minimizing $\mathcal{G}$ we can 
always find a pure state with real wave function for which $\gamma_{xx}[I_i|I_j] 
= \gamma_{pp}[I_i|I_j] = 0$ for $1 \leq i,j \leq k$, $i \not= j$. The notation 
$A[I|I']$, where $I$ and $I'$ are sets of indices, is used to denote the 
submatrix of $A$ formed by the intersection of rows with indices in $I$ and 
columns with indices in $I'$. Using such a state, we can improve the lower 
bound for $\mathcal{G}$.

Due to the relations $\langle\hat{x}_i\hat{x}_j\rangle = 
\langle\hat{p}_i\hat{p}_j\rangle = 0$ for separable $i$ and $j$ we have 
the equality 
\begin{equation}
    \tr(X\gamma_{xx}) + \tr(P\gamma_{pp}) = 
    \tr(X_{\vec{u}}\gamma_{xx}) + \tr(P_{\vec{v}}\gamma_{pp}),
\end{equation}
where $X_{\vec{u}}$ is obtained from $X = (x_{ij})^n_{i,j=1}$ by replacing its 
elements corresponding to zero elements of $\gamma_{xx}$ by arbitrary real 
numbers $u_i$ subject to the condition that $X_{\vec{u}}$ is symmetric and 
positive semidefinite and the same procedure is applied to $P$ to produce 
$P_{\vec{v}}$. In other words, we can replace all elements of the submatrices 
of the form $X[I_i|I_j]$ and $P[I_i|I_j]$, $i \not= j$, by arbitrary real 
numbers in such a way that the resulting matrices are again symmetric and 
positive semidefinite. This construction is better illustrated by an example. 
For $n=4$, consider $2|134$- and $1|2|34$-factorizable states. In the former 
case we have $I_1 = \{2\}$, $I_2 = \{1, 3, 4\}$, and the the latter case we have 
$I_1 = \{1\}$, $I_2 = \{2\}$, $I_3 = \{3, 4\}$. The matrices $X_{\vec{u}}$ 
corresponding to these two cases are
\begin{equation}\label{eq:P}
    \begin{pmatrix}
        x_{11} & \tikzmarkin[hor=style orange]{el 1}u_1\tikzmarkend{el 1} & 
x_{13} & x_{14} \\[1mm]
        \tikzmarkin[hor=style orange]{el 2}u_1\tikzmarkend{el 2} & x_{22} & 
\tikzmarkin[hor=style orange]{el 3} u_2 & u_3\tikzmarkend{el 3} \\[1mm]
        x_{13} & \tikzmarkin[ver=style orange]{col 1} u_2 & x_{33} & x_{34} 
\\[1mm]
        x_{14} & u_3\tikzmarkend{col 1} & x_{34} & x_{44}
    \end{pmatrix}, \ 
    \begin{pmatrix}
        x_{11} & \tikzmarkin[hor=style orange]{el 4}u_1\tikzmarkend{el 4} & 
\tikzmarkin[hor=style cyan]{el 7}u_2 & u_3\tikzmarkend{el 7} \\[1mm]
        \tikzmarkin[ver=style orange]{el 5}u_1\tikzmarkend{el 5} & x_{22} & 
\tikzmarkin[hor=style green]{el 6} u_4 & u_5\tikzmarkend{el 6} \\[1mm]
        \tikzmarkin[ver=style cyan]{el 8}u_2 & \tikzmarkin[ver=style 
green]{col 2} u_4 & x_{33} & x_{34} 
\\[1mm]
        u_3\tikzmarkend{el 8} & u_5\tikzmarkend{col 2} & x_{34} & x_{44}
    \end{pmatrix}.
\end{equation}
In the first case we replace elements of the submatrices $X[I_1|I_2]$ and 
$X[I_2|I_1]$ by arbitrary numbers, and in the second case we replace 
submatrices $X[I_1|I_2]$, $X[I_2|I_1]$, $X[I_1|I_3]$, $X[I_3|I_1]$, 
$X[I_2|I_3]$, $X[I_3|I_2]$. Different submatrices are marked by different 
colors (symmetric parts are marked by the same color). The more factorizable 
parts the state has the more elements can be replaced by arbitrary numbers. For 
a completely factorizable state we can freely choose all the off-diagonal 
entries. 

In order not to overload the notation, we fix some kind of separability, i.e., 
some decomposition $\mathcal{I} = \{I_1, \ldots, I_k\}$ of the indices, and use 
it in all the considerations below. Applying the inequality \eqref{eq:minG3}, 
we find that for a pure $\mathcal{I}$-factorizable state with real wave 
function, and thus for all $\mathcal{I}$-separable states, we have
\begin{equation}\label{eq:minGsep}
    \mathcal{G}(X, P) \geq \mathcal{B}_{\mathcal{I}}(X, P) = 
    \max_{\vec{u},\vec{v}} \tr \sqrt{\sqrt{X_{\vec{u}}} P_{\vec{v}} 
    \sqrt{X_{\vec{u}}}},
\end{equation}
where the optimization is over the points $\vec{u}$ and $\vec{v}$ such that 
$X_{\vec{u}}$ and $P_{\vec{v}}$ are positive definite. For example, for a 
bipartition $\{I_1, I_2\}$ with $|I_1| = m$, $|I_2| = n-m$ the vectors 
$\vec{u}$ 
and $\vec{v}$ have $m(n-m)$ components, so in this case the optimization 
problem 
has $2m(n-m)$ variables. The full separability optimization problem has 
$n(n-1)$ components. 

Note that for the vectors $\vec{u}_0$ and $\vec{v}_0$ with the 
corresponding elements of the matrices $X$ and $P$, respectively, we have 
$X_{\vec{u}_0} = X$ and $P_{\vec{v}_0} = P$ and thus for any partition
$\mathcal{B}_{\mathcal{I}}(X, P) \geq \mathcal{B}(X, P)$. This inequality 
is strict for most of the matrices $X$ and $P$. It follows that the inequality 
\eqref{eq:minGsep} gives a stronger bound on $\mathcal{G}$ then does the 
inequality \eqref{eq:minG3}. Moreover, in the same way we obtain that if 
$\mathcal{I}'$ is finer than $\mathcal{I}$ then $\mathcal{B}_{\mathcal{I}'}(X, 
P) \geq \mathcal{B}_{\mathcal{I}}(X, P)$, since in this case we have more 
variables to optimize over, and thus the condition for $\mathcal{I}'$-separable 
states gives a stronger bound then the condition for 
$\mathcal{I}$-separability. 
The condition for full separability gives the strongest bound of all. 
We have just obtained a hierarchy of separability conditions that 
mirrors the partition lattice in Fig.~\ref{fig:partitions}.

To each kind of separability $\mathcal{I}$ corresponds its own maximization 
problem of its own dimension. There are $2^{n-1}-1$ different bipartitions of 
the indices of an $n$-partite state and many more partitions into three or more 
parts. If for a given state there is a pair of matrices $X$ and $P$ such that 
an inequality \eqref{eq:minGsep} is violated, then the state is not 
$\mathcal{I}$-separable. If there are $X$ and $P$ such that the inequalities 
\eqref{eq:minGsep} are violated for all bipartitions simultaneously, then the 
state is genuine multipartite entangled.

\section{Rank-one matrices} 
\label{sec:r1}

We now show that the results of Ref.~\cite{PhysRevA.67.052315} are just a 
special case of the general 
inequalities \eqref{eq:minGsep} and give an analytical solution of these 
optimization problems in this special case. Consider the rank-one 
matrices $X = \vec{h}\vec{h}^{\mathrm{T}}$ and $P = 
\vec{g}\vec{g}^{\mathrm{T}}$. The square root of a rank-one matrix $A = 
\vec{a}\vec{a}^{\mathrm{T}}$ is given by $\sqrt{A} = 
\vec{a}\vec{a}^{\mathrm{T}}/\|\vec{a}\|$, so we have 
$\mathcal{B}(X, P) = |(\vec{h}, \vec{g})|$. As a concrete example let us 
consider four-partite case and $\mathcal{I}=1|2|34$-separable states. We are 
free to change some elements of the matrices $X$ and $P$. Let us just change 
the sign of the $X$ and $P$'s elements that are marked in Eq.~\eqref{eq:P}:
\begin{equation}
    X' = 
    \begin{pmatrix}
        h^2_1 & \tikzmarkin[hor=style orange]{el 
101}(0.08,-0.12)(-0.08,0.27)\pm h_1 h_2\tikzmarkend{el 101} & 
\tikzmarkin[hor=style cyan]{el 107}(0.08,-0.12)(-0.08,0.27)\pm h_1 h_3 & \pm 
h_1 h_4\tikzmarkend{el 107} \\[1mm]
        \tikzmarkin[hor=style orange]{el 102}(0.08,-0.12)(-0.08,0.27)\pm h_1 
h_2\tikzmarkend{el 102} & h^2_2 & \tikzmarkin[hor=style green]{el 
106}(0.08,-0.12)(-0.08,0.27)\pm h_2 h_3 & \pm h_2 h_4\tikzmarkend{el 106} 
\\[1mm]
        \tikzmarkin[ver=style cyan]{el 105}(0.14,-0.12)(-0.08,0.27)\pm h_1 h_3 
& \tikzmarkin[ver=style green]{el 108}(0.14,-0.12)(-0.08,0.27)\pm h_2 h_3 & 
h^2_3 & h_3 h_4 \\[1mm]
        \pm h_1 h_4\tikzmarkend{el 105} & \pm h_2 h_4\tikzmarkend{el 108} & 
h_3 h_4 & h^2_4
    \end{pmatrix}.
\end{equation}
For appropriate combinations of signs we can get that $X' = 
\vec{h}'\vec{h}^{\prime\mathrm{T}}$ and $P' = 
\vec{g}'\vec{g}^{\prime\mathrm{T}}$, where the new vectors read as $\vec{h}' = 
(\pm h_1, \pm h_2, h_3, h_4)$ and $\vec{g}' = (\pm g_1, \pm g_2, g_3, g_4)$, 
and thus 
\begin{equation}
\begin{split}
    &\mathcal{B}_{\mathcal{I}}(X, P) \geq \max |(\vec{h}', \vec{g}')| \\
    &= \max |\pm h_1 g_1 \pm h_2 g_2 + h_3 g_3 + h_4 g_4| \\
    &= |h_1 g_1| + |h_2 g_2| + |h_3 g_3 + h_4 g_4| \geq \mathcal{B}(X, P).
\end{split}
\end{equation}
This result can be extended to 
all $n$ and arbitrary kind of separability and coincides with the results 
obtained in Ref.~\cite{PhysRevA.67.052315}.

Note that the inequality $\vec{h}^{\mathrm{T}} \gamma_{xx} \vec{h} + 
\vec{g}^{\mathrm{T}} \gamma_{pp} \vec{g} \geq |(\vec{h}, \vec{g})|$ is 
equivalent to the inequalities
\begin{equation}
    \begin{pmatrix}
        \gamma_{xx} & \frac{1}{2}E \\[1mm]
        \frac{1}{2}E & \gamma_{pp}
    \end{pmatrix} 
    \geq 0.
\end{equation}
The same idea can be applied to the separability conditions and we arrive to 
the conditions for $\mathcal{I}$-separable states
\begin{equation}\label{eq:hg}
    \begin{pmatrix}
        \gamma_{xx} & \frac{1}{2} E_{\mathcal{I}} \\[1mm]
        \frac{1}{2} E_{\mathcal{I}} & \gamma_{pp}
    \end{pmatrix}
    \geq 0,
\end{equation}
where $E_{\mathcal{I}}$ is a diagonal matrix with diagonal elements equal to 
$\pm 1$ so that the elements with indices in the same group $I_j$ have the 
same sign. These conditions work in many cases, but sometimes the more general 
conditions are needed. In the next section we give an example of a PPT state 
that satisfies the inequalities \eqref{eq:hg} but violates the general 
inequalities \eqref{eq:minGsep}.

\section{An example of PPT state}
\label{sec:ppt}

Consider a four-partite state with the covariance matrix given by 
\cite{PhysRevLett.86.3658, NewJPhys.8.51}
\begin{equation}\label{eq:s4}
\begin{split}
    \gamma_{xx} &= \frac{1}{2}
    \begin{pmatrix}
        2 & 0 & 1 & 0 \\[1mm]
        0 & 2 & 0 & -1 \\[1mm]
        1 & 0 & 2 & 0 \\[1mm]
        0 & -1 & 0 & 2
    \end{pmatrix}, \\
    \gamma_{pp} &= \frac{1}{2}
    \begin{pmatrix}
        1 & 0 & 0 & -1 \\[1mm]
        0 & 1 & -1 & 0 \\[1mm]
        0 & -1 & 4 & 0 \\[1mm]
        -1 & 0 & 0 & 4
    \end{pmatrix}.
\end{split}
\end{equation}
The matrix $\gamma = \left(\begin{smallmatrix} \gamma_{xx} & 0 \\ 0 
& \gamma_{pp} \end{smallmatrix}\right)$ is a covariance matrix of a quantum 
state, since $\gamma + (i/2)J \geq 0$. Partial transpositions of $(1, 
3)$-kinds (that is 
transpositions $1|234$, $2|134$, $3|124$ and $4|123$) are negative, and these 
negativities are easily detected by the conditions \eqref{eq:hg}. For 
example, the matrix $E_{1|234}$ reads as
\begin{equation}
    E_{1|234} = 
    \begin{pmatrix}
        -1 & 0 & 0 & 0 \\[1mm]
        0 & 1 & 0 & 0 \\[1mm]
        0 & 0 & 1 & 0 \\[1mm]
        0 & 0 & 0 & 1
    \end{pmatrix}
\end{equation}
and the matrix
$
    \left(
    \begin{smallmatrix}
        \gamma_{xx} & E_{1|234}/2 \\
        E_{1|234}/2 & \gamma_{pp}
    \end{smallmatrix}
    \right)
$
has negative eigenvalues. On the other hand, the partial transpositions of $(2, 
2)$-kinds are all positive and the matrices \eqref{eq:hg} with
$\mathcal{I} = 12|34, 13|24, 14|23$ are positive semidefinite, so the 
test \eqref{eq:hg} does not detect any entanglement of a $(2, 2)$-kind. 
Nevertheless, the state \eqref{eq:s4} is entangled in any of this kind.

To demonstrate this, consider the positive semidefinite matrices
\begin{equation}
\begin{split}
    X &= 
    \begin{pmatrix}
        x & 0 & -\sqrt{xy} & 0 \\[1mm]
        0 & x & 0 & \sqrt{xy} \\[1mm]
        -\sqrt{xy} & 0 & y & 0 \\[1mm]
        0 & \sqrt{xy} & 0 & y
    \end{pmatrix} \\
    P &= 
    \begin{pmatrix}
        p & 0 & 0 & \sqrt{pq} \\[1mm]
        0 & p & \sqrt{pq} & 0 \\[1mm]
        0 & \sqrt{pq} & q & 0 \\[1mm]
        \sqrt{pq} & 0 & 0 & q
    \end{pmatrix},
\end{split}
\end{equation}
where $x$, $y$, $p$ and $q$ are some positive numbers to be determined. The 
quantity $\mathcal{G}(X, P) = \tr(X\gamma_{xx}) + \tr(P\gamma_{pp})$ reads as
$\mathcal{G}(X, P) = 2x + 2y -2\sqrt{xy} + p + 4q - 2\sqrt{pq}$. To compute 
$\mathcal{B}_{12|34}(X, P)$, note that 
\begin{equation}
    \mathcal{B}_{12|34}(X, P) \geq \sqrt{\sqrt{X'}P'\sqrt{X'}},
\end{equation}
where the matrices $X'$ and $P'$ are defined to be (since we can play with the 
elements marked by color)
\begin{equation}\label{eq:X'P'}
    X' = 
    \begin{pmatrix}
        x & 0 & \tikzmarkin[hor=style orange]{m1} 0 & 0 \\[1mm]
        0 & x & 0 & 0 \tikzmarkend{m1}\\[1mm]
        \tikzmarkin[hor=style orange]{m2} 0 & 0 & y & 0 \\[1mm]
        0 & 0 \tikzmarkend{m2} & 0 & y
    \end{pmatrix}, \quad
    P' = 
    \begin{pmatrix}
        p & 0 & \tikzmarkin[hor=style orange]{m3} 0 & 0 \\[1mm]
        0 & p & 0 & 0 \tikzmarkend{m3} \\[1mm]
        \tikzmarkin[hor=style orange]{m4} 0 & 0 & q & 0 \\[1mm]
        0 & 0 \tikzmarkend{m4} & 0 & q
    \end{pmatrix},
\end{equation}
and thus for the boundary $\mathcal{B}_{12|34}$ we have the inequality
\begin{equation}\label{eq:BI}
    \mathcal{B}_{12|34} \geq 2(\sqrt{xp}+\sqrt{yq}).
\end{equation}
It is easy to check that for the following values of the parameters:
\begin{equation}
\begin{split}
    x &= 0.144375, \quad y = 0.084087, \\
    p &= 0.232000, \quad q = 0.039543
\end{split}
\end{equation}
we have $\mathcal{G}(X, P) = 0.435170$, while according to Eq.~\eqref{eq:BI} we 
have $\mathcal{B}_{12|34} \geq 0.481359$. We see that $\mathcal{B}_{12|34} > 
\mathcal{G}(X, P)$, and the PPT state Eq.~\eqref{eq:s4} is $12|34$-entangled. 
The construction for the partitions $13|24$ and $14|23$ is similar.

\section{Genuine entanglement condition}
\label{sec:ge}

The inequalities \eqref{eq:minGsep} allow one to test states for entanglement 
of some kind, but the number of these kinds grows extremely fast with the 
number of parts. We now derive an analytical condition for genuine multipartite 
entanglement. It is a \textit{single} condition that does not require testing 
exponentially many bipartitions, however, it does not provide the best possible 
bound. Consider the quantity
\begin{equation}\label{eq:Gn}
    \mathcal{G}_n = \sum_{1 \leq i < j \leq n} 
    \langle (\hat{x}_i + \hat{x}_j)^2 + (\hat{p}_i - \hat{p}_j)^2 \rangle.
\end{equation}
It is the general quantity $\mathcal{G}$ with the following matrices $X = X_n$ 
and $P = P_n$:
\begin{equation}
\begin{split}
    X_n = 
    \begin{pmatrix}
        n-1 & 1 & \ldots & 1 & 1 \\[1mm]
        1 & n-1 & \ldots & 1 & 1 \\[1mm]
        \hdotsfor{5} \\[1mm]
        1 & 1 & \ldots & n-1 & 1 \\[1mm]
        1 & 1 & \ldots & 1 & n-1 \\
    \end{pmatrix} \\
    P_n = 
    \begin{pmatrix}
        n-1 & -1 & \ldots & -1 & -1 \\[1mm]
        -1 & n-1 & \ldots & -1 & -1 \\[1mm]
        \hdotsfor{5} \\[1mm]
        -1 & -1 & \ldots & n-1 & -1 \\[1mm]
        -1 & -1 & \ldots & -1 & n-1 \\
    \end{pmatrix}
\end{split}
\end{equation}
Since matrices $X_n$ and $P_n$ are completely symmetric with 
respect to different parts, it is enough to consider only bipartitions of the 
form $1, \ldots, k | k+1, \ldots, n$ for $k \leq n/2$. We do not 
change the elements of $X_n$, and in the matrix $P_n$ we set all $v_i$ to $1$ 
(so we change the matrix elements from $-1$ to $1$). Denote the resulting 
matrix by $P'_{n, k}$. The matrices $X_n$ and $P'_{n, k}$ commute, so that $\tr 
(\sqrt{X_n} P'_{n, k} \sqrt{X_n})^{1/2}$ is easy to compute 
\begin{equation}
    \tr\sqrt{X_n P'_{n, k}} = (n-1)\sqrt{n(n-2)} + 4\beta 
    \frac{k(n-k)}{\sqrt{n}}, 
\end{equation}
where the number $\beta$ is given by
\begin{equation}
    \beta = \frac{1}{\sqrt{2n-2} + \sqrt{n-2}}. 
\end{equation}
The minimum of this expression over $1 \leq k \leq n/2$ is attained for $k=1$. 
We have just obtained the following result: Any biseparable state must satisfy 
the inequality 
\begin{equation}\label{eq:Gg}
    \mathcal{G}_n \geq (n-1)\sqrt{n(n-2)} + \frac{4(n-1)}{\sqrt{n}(\sqrt{2n-2} 
    + \sqrt{n-2})}.
\end{equation}
If this inequality is violated, then the state is genuine multipartite 
entangled. Table~\ref{tbl:sep} summarizes the lower bounds of $\mathcal{G}_n$ 
for some $n$ obtained with the analytical condition \eqref{eq:Gg} and computed 
numerically from Eq.~\eqref{eq:minGsep}.

\begin{table}[ht]
\begin{tabular}{l|D{.}{.}{2.2}D{.}{.}{2.2} 
D{.}{.}{4.2}D{.}{.}{4.2}D{.}{.}{4.2}D{.}{.}{4.2}D{.}{.}{4.2}D{.}{.}{4.2}}
\toprule[0.6pt]
 $n$ & 2 & \multicolumn{1}{c}{\hfil 3} & \multicolumn{1}{c}{\hfil 4} & 
\multicolumn{1}{c}{\hfil 5} & \multicolumn{1}{c}{\hfil 6} & 
\multicolumn{1}{c}{\hfil 7} & \multicolumn{1}{c}{\hfil 8} \\
\midrule[0.4pt]\\[-2mm]
q & 0 & 3.46 & 8.48 & 15.49 & 24.49 & 35.49 & 48.49 \\[1mm]
a & 2 & 5.00 & 10.03 & 17.06 & 26.07 & 37.08 & 50.09 \\[1mm]
b & 2 & 5.46 & 10.89 & 18.26 & 27.59 & 38.89 & 52.17 \\[1mm]
f & 2 & 6 & 12 & 20 & 30 & 42 & 56 \\[1mm]
\bottomrule[0.6pt]
\end{tabular}
\caption{Lower bounds for $\mathcal{G}_n$ for different kinds of 
$n$-partite states. The \textit{q} row is quantumness bound, 
$(n-1)\sqrt{n(n-2)}$. The \textit{a} row is the biseparability 
bound given by the analytical expression \eqref{eq:Gg}. The \textit{b} row is 
the true biseparability bound given by the solution of the optimization problem 
\eqref{eq:minGsep}. The last row, \textit{f}, is the full separability bound, 
$n(n-1)$.} \label{tbl:sep}
\end{table}
A similar genuine entanglement condition has been obtained in 
Ref.~\cite{PhysRevA.67.052315} in terms of rank-one matrices. The gap between 
quantum bound and biseparability bound there decreases as $O(1/n)$, where $n$ 
is the number of parts\footnote{In fact, in Ref.~\cite{PhysRevA.67.052315} 
the term \textit{genuine entanglement} was used in a different, weaker, 
sense. But the condition obtained there happens to work for genuine 
entanglement in the present, stronger, sense.}. In our condition this gap is 
$O(1)$ and so it does not tend to zero for large $n$.

\section{Convex optimization and experimental errors}
\label{sec:co}

To demonstrate entanglement of a kind $\mathcal{I}$ we need to find a pair of 
matrices $X$ and $P$ that violate the condition \eqref{eq:minGsep}. In practice, 
however, we should take into account the errors in the measurement of the matrix 
elements of $\gamma_{xx}$ and $\gamma_{pp}$. Assuming that the errors in the 
individual elements of $\gamma_{xx}$ and $\gamma_{pp}$ are independent, the 
standard deviation $\sigma(X, P)$ of $\mathcal{G}(X, P)$ is given by the 
expression 
\begin{equation}
    \sigma^2(X, P) = \sum^n_{i, j=1} (x^2_{ij}\sigma^2_{xx, ij} + 
p^2_{ij}\sigma^2_{pp, ij}),
\end{equation}
where $\sigma_{xx, ij}$ and $\sigma_{pp, ij}$ are the standard deviations of 
individual elements of $\gamma_{xx}$ and $\gamma_{pp}$ respectively. So, to be 
on the safe side the right inequality reads as
\begin{equation}\label{eq:E}
    \mathcal{E}_{\mathcal{I}}(X,P) = \mathcal{G}(X, P) + s\,\sigma(X, P)
    -\mathcal{B}_{\mathcal{I}}(X, P) \geq 0,
\end{equation}
where $s$ is the level of certainty with which we can claim that the state is 
entangled. We prove that the function $\mathcal{E}_{\mathcal{I}}(X,P)$ is 
convex on the set of all pairs of semidefinite matrices $(X,P)$. 

To prove the convexity of $\mathcal{E}_{\mathcal{I}}(X,P)$ we have to prove the 
concavity of $\mathcal{B}_{\mathcal{I}}(X, P)$ because the convexity of the 
other two terms of $\mathcal{E}_{\mathcal{I}}(X,P)$ is obvious. The key element 
of this proof is the fact that $\tr \sqrt{\sqrt{X_{\vec{u}}} P_{\vec{v}} 
\sqrt{X_{\vec{u}}}}$ is jointly concave with respect to all four variables $X$, 
$P$, $\vec{u}$ and $\vec{v}$. Due to the equality 
\begin{equation}
    \tr \sqrt{\sqrt{X} P \sqrt{X}} = \min_{\gamma_{xx},\gamma_{pp}} (\tr(X 
\gamma_{xx}) + \tr(P \gamma_{pp}))
\end{equation}
it immediately follows $\tr \sqrt{\sqrt{X} P \sqrt{X}}$ is concave with 
respect to $X$ and $P$ as the minimum of a family of concave (in fact, linear) 
functions. From the relation
\begin{equation}
\begin{split}
    (\theta X_1 &+ (1-\theta)X_2)'(\theta\vec{u}_1 + (1-\theta)\vec{u}_2) \\
    &= \theta X'_1(\vec{u}_1) + (1-\theta)X'_2(\vec{u}_2),
\end{split}
\end{equation}
$0 \leq \theta \leq 1$, and similar relation for $P$ we derive the 
joint concavity of $\tr \sqrt{\sqrt{X_{\vec{u}}} P_{\vec{v}} 
\sqrt{X_{\vec{u}}}}$. 

We now have three sets --- the set $\Omega_{x,1} \times \Omega_{p,1}$ of points 
$(\vec{u},\vec{v})$ where $X'_1(\vec{u})$ and $P'_1(\vec{v})$ are positive 
definite, the similar set $\Omega_{x,2} \times \Omega_{p,2}$ for $X_2$ and 
$P_2$, and the set $\Omega_{x} \times \Omega_{p}$ for $X = \theta X_1 + 
(1-\theta) X_2$ and $P = \theta P_1 + (1-\theta) P_2$. In general, these are 
distinct sets, but one can easily see that $\theta \Omega_{x,1} \times 
\Omega_{p,1} + (1-\theta)\Omega_{x,2} \times \Omega_{p,2} \subseteq \Omega_{x} 
\times \Omega_{p}$. The standard argument given in Ref.~\cite{convex-opt} can be 
applied here to conclude that $\mathcal{B}_{\mathcal{I}}(X, P)$ is concave 
as the maximum of a jointly concave function over a convex set. This finishes 
the proof of the convexity of the function $\mathcal{E}_{\mathcal{I}}(X, P)$ 
defined by Eq.~\eqref{eq:E}.

To violate the inequality \eqref{eq:E} we thus have to optimize 
$\mathcal{E}_{\mathcal{I}}(X,P)$ over $X$ and $P$ and check whether the optimal 
value is negative or not. This optimization problem has $n(n+1)$ variables, the 
elements of $X$ and $P$. Since the function $\mathcal{E}_{\mathcal{I}}(X, P)$ is 
homogeneous, $\mathcal{E}_{\mathcal{I}}(\lambda X, \lambda P) = 
\lambda \mathcal{E}_{\mathcal{I}}(X, P)$ for $\lambda \geq 0$, it makes sense 
to put some condition on the matrices $X$ and $P$. The simplest is a linear 
condition, for example, the condition $\tr(X\gamma_{xx}+P\gamma_{pp}) = C$, 
where $C$ is an arbitrary fixed positive constant. We thus arrive to the 
following $\mathcal{I}$-separability condition:
\begin{equation}\label{eq:O}
    \min_{\tr(X\gamma_{xx}+P\gamma_{pp}) = C} (s\,\sigma(X,P) - 
    \mathcal{B}_{\mathcal{I}}(X, P) \geq -C. 
\end{equation}
If, for given $\gamma_{xx}$, $\gamma_{pp}$, $\sigma_{xx}$ and $\sigma_{pp}$, 
this inequality is violated, i.e., if this minimum drops below $-C$ then the 
state in question is not separable of the corresponding kind. If these 
inequalities can be violated for all bipartitions simultaneously (by the same 
pair of matrices $X$ and $P$), then the state is genuine multipartite 
entangled. The methods to solve convex optimization problems like the one given 
by Eq.~\eqref{eq:O} are discussed, e.g, in Ref.~\cite{convex-opt}. 

What $s$ should we choose? Usually, the "three-sigma rule", $s=3$ is 
applied \cite{lnm}. To better understand what values of $s$ in Eq.~\eqref{eq:E} 
are sufficient to guarantee that our results are correct we need to know the 
probability that the result of a measurement lies outside $s$ sigma interval. 
For a Gaussian probability distribution with the mean $\mu$ and the standard 
deviation $\sigma$ this probability is given by the expression
\begin{equation}\label{eq:Ps}
    \mathbf{P}(s) = 1 - \int^{\mu + s \sigma}_{\mu - s \sigma} 
\frac{1}{\sqrt{2\pi}\sigma} e^{-\frac{(x-\mu)^2}{2\sigma^2}} \, dx = 
1-\erf\left(\frac{s}{\sqrt{2}}\right),
\end{equation}
which depends only on $s$. This probability decreases very quickly as $s$ 
growth. The order of values of $\mathbf{P}(s)$ for some $s$ are shown in the 
table below.
\begin{table}[ht]
\renewcommand{\arraystretch}{1.5}
\setlength{\tabcolsep}{4pt}
\begin{tabular}{c|llllllll}
 $s$ & \hfil 1 & \hfil 2 & \hfil 3 & \hfil 4 & \hfil 5 & \hfil 6 & \hfil 7 & 
\hfil 10 \\ 
 $\mathbf{P}(s)$ & $0.32$ & $0.05$ & $10^{-3}$ & $10^{-4}$ & $10^{-6}$ & 
$10^{-9}$ & $10^{-12}$ & $10^{-23}$
\end{tabular}
\end{table}
In many cases, the value of $s=3$ is sufficient (the three-sigma 
rule). For $s \geq 5$ this probability is negligible, and for $s \geq 6$ it is 
practically zero. Even if the real probability distribution is not perfectly 
Gaussian it is unlikely to have long tail, so Eq.~\eqref{eq:Ps} gives a 
reasonable estimate. Even if this estimate is wrong by several orders of 
magnitude, provided that we have verified violation with $s \geq 6$ we are 
still on the safe side. The larger $s$ we set the larger the probability of the 
correct result is, but the more difficult it will be to find a violation with 
such $s$. From this table we can conclude that we should search for a violation 
with $s$ not smaller than 3 and not larger than 6 --- the event of getting the 
right result outside of six sigma interval is practically impossible.

We see that if we can violate our inequalities with $s=6$ then it practically 
guarantees that our conclusion is correct. In Appendix B we demonstrate the 
usefulness of our approach by applying to some four-, six- and ten-partite 
realistic states. We also demonstrate that the four-partite state in question is 
genuine multipartite entangled.

\section{Conclusion} 

We have developed a method to test continuous-variable multipartite states for 
arbitrary kinds of entanglement. Our approach allows both numerical and 
analytical treatment. Numerically, it reduces to a convex optimization problem, 
which allows fast and accurate solution. We have shown that it is very efficient 
at detecting ordinary entanglement and can detect genuine multipartite 
entanglement in a reasonable amount of time. Analytically, it allows to 
reproduce (and thus generalize) some known results as well as to obtain an 
analytical genuine multipartite entanglement condition. With our approach we can 
easily obtain a trace-class inequality, which is difficult to get in a direct 
way.

\appendix

\section{Trace inequalities}

If matrices $X$ and $P$ commute then the right-hand side of 
Eq.~\eqref{eq:minG2} reduces to $\tr \sqrt{X}\sqrt{P} = \tr 
\sqrt{P}\sqrt{X}$. This expression, $\tr(\sqrt{X}\sqrt{P})$, is a lower bound 
for $\mathcal{G}$ independent of commutation properties of $X$ and $P$. For a 
real wave function $f(\vec{x})$ we have the equality 
\begin{equation}
    \mathcal{G}(X, P) = \int (\vec{u}^{\mathrm{T}}(\vec{x}) X \vec{u}(\vec{x}) 
    + \vec{v}^{\mathrm{T}}(\vec{x}) P \vec{v}(\vec{x})) \, d\vec{x},
\end{equation}
where the 
vector fields $\vec{u}(\vec{x})$ and $\vec{v}(\vec{x})$ are 
defined via $\vec{u}(\vec{x}) = f(\vec{x})\vec{x}$ and $\vec{v}(\vec{x}) 
= \nabla f(\vec{x})$. We can write this equality in a more compact form as
\begin{equation}
    \mathcal{G}(X, P) = \int (\|\tilde{\vec{u}}(\vec{x})\|^2 + 
    \|\tilde{\vec{v}}(\vec{x})\|^2) \, d\vec{x},    
\end{equation}
where the new vector fields are defined via $\tilde{\vec{u}} = \sqrt{X} 
\vec{u}$ 
and $\tilde{\vec{v}} = \sqrt{P} \vec{v}$. Now we can estimate $\mathcal{G}$ as 
follows: 
\begin{equation}
\begin{split}
    \mathcal{G}(X, P) &\geq 2| \int (\tilde{\vec{u}}(\vec{x}), 
\tilde{\vec{v}}(\vec{x})) \, d\vec{x}| \\
    &= 2\left| \int \vec{x}^{\mathrm{T}} \sqrt{X}\sqrt{P} 
(\nabla f)(\vec{x}) f(\vec{x}) \, d\vec{x} \right|.
\end{split}
\end{equation}
From the relation 
\begin{equation}
    \int u_j(\vec{x}) v_k(\vec{x}) \, d\vec{x} = \int x_j f(\vec{x})
    \frac{\partial f(\vec{x})}{\partial x_k} \, d\vec{x} = 
    -\frac{1}{2}\delta_{jk},
\end{equation}
we get the inequality $\mathcal{G}(X, P) \geq \tr(\sqrt{X}\sqrt{P})$.

We thus have two lower bounds for $\mathcal{G}$ --- the tight one is given by 
the inequality \eqref{eq:minG2} and the other one, not necessarily tight, 
have just been obtained with the help of Cauchy-Schwarz inequality. Since the 
tight bound is the best bound possible, we derive the following inequality for a 
pair of positive definite matrices $X$ and $P$:
\begin{equation}\label{eq:XP}
    \tr \sqrt{\sqrt{X}P\sqrt{X}} \geq \tr(\sqrt{X}\sqrt{P}) = 
    \tr (\sqrt[4]{X}\sqrt{P}\sqrt[4]{X}).
\end{equation}
This inequality is a special case of Araki-Lieb-Thirring trace inequalities 
\cite{PhysRevLett.35.687, lieb-thirring, Lett.Math.Phys.19.167}, which also 
have quantum mechanical background and read as 
\begin{equation}
    \tr (A^{1/2}BA^{1/2})^{rq} \geq \tr (A^{r/2} B^r A^{r/2})^q,
\end{equation}
where $A$ and $B$ are arbitrary positive definite 
matrices, $q \geq 0$ and $0 \leq r \leq 1$. The case of $q=1$ and $r = 1/2$ 
corresponds to the inequality \eqref{eq:XP}.

\section{Application to realistic matrices}

It happens that rank-one matrices work surprisingly well. Consider the 
four-partite state that was analyzed in Ref.~\cite{PhysRevLett.114.050501}. It 
has the following covariance matrix:
\begin{equation}\label{eq:cov4}
\begin{split}
    \gamma_{xx} &= 
    \begin{pmatrix}
        1.09921 & 0.16092 & -0.17609 & -0.84831 \\[1mm]
        0.16092 & 0.40938 & -0.16060 & -0.18963 \\[1mm]
        -0.17609 & -0.16060 & 0.46060 & 0.04319 \\[1mm]
        -0.84831 & -0.18963 & 0.04319 & 1.06419
    \end{pmatrix} \\
    \gamma_{pp} &= 
    \begin{pmatrix}
        1.09921 & 0.35533 & 0.36439 & 0.91386 \\[1mm]
        0.35533 & 0.92282 & 0.57440 & 0.43388 \\[1mm]
        0.36439 & 0.57440 & 1.04339 & 0.34868 \\[1mm]
        0.91386 & 0.43388 & 0.34868 & 1.06419
    \end{pmatrix} 
\end{split}
\end{equation}
The standard deviation matrix reads as
\begin{equation}\label{eq:sig4}
\begin{split}
    \sigma_{xx} &= 
    \begin{pmatrix}
        0.00327 & 0.01041 & 0.00894 & 0.00647 \\[1mm]
        0.01041 & 0.00822 & 0.01848 & 0.01899 \\[1mm]
        0.00894 & 0.01848 & 0.00861 & 0.01345 \\[1mm]
        0.00647 & 0.01899 & 0.01345 & 0.00549
    \end{pmatrix} \\
    \sigma_{pp} &= 
    \begin{pmatrix}
        0.00458 & 0.01009 & 0.02767 & 0.04289 \\[1mm]
        0.01009 & 0.01023 & 0.02101 & 0.02085 \\[1mm]
        0.02767 & 0.02101 & 0.01466 & 0.01955 \\[1mm]
        0.04289 & 0.02085 & 0.01955 & 0.00455
    \end{pmatrix} 
\end{split}
\end{equation}
A very simple way to search for violation of the condition \eqref{eq:E} is 
to randomly generate 4-vectors $\vec{h}$ and $\vec{g}$ and check whether this 
condition is violated by the rank-one matrices $X = 
\vec{h}\vec{h}^{\mathrm{T}}$ and $P = \vec{g}\vec{g}^{\mathrm{T}}$, and if it 
is, how strong the violation is. Then just record the maximal observed 
violation. As a measure of violation we use the quantity 
\begin{equation}
    s = \frac{\mathcal{B}_{\mathcal{I}}(X, P) - \mathcal{G}(X, P)}{\sigma(X, 
P)}.
\end{equation}
This approach requires only simple matrix algebra manipulations, which can be 
done very efficiently with tools like Intel Math Kernel Library. A simple 
parallel Fortran program has been written and run on a low-end 4-core desktop 
PC. The total time to test all seven possible bipartitions in this four-partite 
case is 4 minutes (using all four cores available). Table~\ref{tbl:4} compares 
our results with those obtained in Ref.~\cite{PhysRevLett.114.050501}. We see 
that for the state under study our approach is superior to that of 
Ref.~\cite{PhysRevLett.114.050501} (which uses a genetic algorithm to find the 
best violation), since it is simpler and gives better results, though, as we 
have mentioned before, from practical point of view all violations larger than 
6 are of the same value.

\begin{table}[ht]
\begin{tabular}{|c|D{.}{.}{3.3}|D{.}{.}{3.3}|l|}
\toprule[0.6pt]
 \multirow{2}{*}{Bipart.} & \multicolumn{2}{c|}{Violation} & 
\multicolumn{1}{c|}{$\vec{h}$} \\ \cline{2-3}
 & 
\multicolumn{1}{c|}{\textrm{Ref.~\cite{PhysRevLett.114.050501}}} & 
\multicolumn{1}{c|}{4m} & \multicolumn{1}{c|}{$\vec{g}$} \\
\midrule[0.4pt]
\multirow{2}{*}{$1|234$} & \multirow{2}{3.5mm}{20.93} & 
\multirow{2}{3.5mm}{26.48} &  $(1.97, -0.01, 0.49, 1.88)$ \\
 & & & $(1.14, 0.18, -0.20, -1.03)$ \\ \hline
\multirow{2}{*}{$2|134$} & \multirow{2}{3.5mm}{13.17} & 
\multirow{2}{3.5mm}{18.08} & $(-0.40, -1.99, -1.48, -0.74)$ \\
 & & & $(-0.15, -1.80, 1.24, 0.68)$ \\ \hline
\multirow{2}{*}{$3|124$} & \multirow{2}{3.5mm}{11.21} & 
\multirow{2}{3.5mm}{16.10} & $(0.31, 1.93, 1.62, 0.32)$ \\
 & & & $(0.23, 1.65, -1.46, -0.19)$ \\ \hline 
\multirow{2}{*}{$4|123$} & \multirow{2}{3.5mm}{21.06} & 
\multirow{2}{3.5mm}{26.57} & $(1.77, 0.18, 0.44, 1.82)$ \\
 & & & $(-0.96, -0.18, -0.12, 1.21)$ \\ \hline
\multirow{2}{*}{$12|34$} & \multirow{2}{3.5mm}{24.34} & 
\multirow{2}{3.5mm}{27.79} & $(1.91, 0.42, 0.75, 1.90)$ \\
 & & & $(-1.06, -0.60, 0.50, 1.19)$ \\ \hline
\multirow{2}{*}{$13|24$} & \multirow{2}{3.5mm}{23.52} & 
\multirow{2}{3.5mm}{26.17} & $(-1.52, -0.17, -0.54, -1.58)$ \\
 & & & $(0.78, -0.09, 0.24, -0.97)$ \\ \hline 
\multirow{2}{*}{$14|23$} & \multirow{2}{2mm}{4.66}  & 
\multirow{2}{2mm}{9.72}  & $(0.22, 1.65, 0.18, 0.77)$ \\
 & & & $(0.28, 1.34, -0.37, -0.95)$ \\
\bottomrule[0.6pt]
\end{tabular}
\caption{The comparison of the violation of the separability condition 
for the state given by the covariance matrix \eqref{eq:cov4} with measurements 
errors given by Eq.~\eqref{eq:sig4}. The second column shows the results 
obtained in Ref.~\cite{PhysRevLett.114.050501}, the third column lists the 
results obtained by randomly testing the inequalities \eqref{eq:E}. The 
total time to perform all seven tests is 4 minutes.}\label{tbl:4}
\end{table}

We now apply our technique to the six-partite state also considered in 
Ref.~\cite{PhysRevLett.114.050501}. We have performed two runs of our program 
on the same hardware as in the previous case, one with a smaller number of 
random trials and the other with 200 times more trials. The first run takes 
approximately 4 minutes to perform all 31 tests, the other one takes around 12 
hours. As Table.~\ref{tbl:6} demonstrates, in this case the optimization based 
on a genetic algorithm gives somewhat larger violations. On the other hand, we 
do not know what computational resources were used to perform that optimization 
and how much time it took. As we have already said, all violation larger than 6 
are of the same practical value and our method produced much better violations 
in just a few minutes on a low-end PC. 

\begin{table}[ht]
\begin{tabular}{|c|D{.}{.}{4.5}|D{.}{.}{4.5}|D{.}{.}{4.5}|}
\toprule[0.6pt]
 \multirow{2}{*}{Bipartition} & \multicolumn{3}{c|}{Violation} \\ \cline{2-4}
 & \multicolumn{1}{c|}{\textrm{Ref.~\cite{PhysRevLett.114.050501}}} & 
\multicolumn{1}{c|}{12h} & \multicolumn{1}{c|}{4m} \\
\midrule[0.4pt]
$1|23456$ & 40.086 & 40.111 & 37.500 \\[1mm]
$2|13456$ & 36.185 & 34.097 & 33.967 \\[1mm]
$3|12456$ & 20.274 & 19.715 & 17.683 \\[1mm]
$4|12356$ & 20.010 & 18.869 & 16.953 \\[1mm]
$5|12346$ & 27.146 & 26.680 & 22.979 \\[1mm]
$6|12345$ & 49.220 & 48.077 & 44.187 \\[1mm]
$12|3456$ & 53.541 & 53.085 & 50.142 \\[1mm]
$13|2456$ & 45.569 & 45.958 & 41.684 \\[1mm]
$14|2356$ & 44.789 & 42.163 & 42.509 \\[1mm]
$15|2346$ & 45.282 & 40.410 & 38.565 \\[1mm]
$16|2345$ & 31.177 & 29.995 & 25.686 \\[1mm]
$23|1456$ & 40.158 & 38.636 & 37.462 \\[1mm]
$24|1356$ & 37.698 & 36.633 & 31.716 \\[1mm]
$25|1346$ & 35.256 & 31.106 & 29.877 \\[1mm]
$26|1345$ & 47.016 & 42.269 & 40.199 \\[1mm]
$34|1256$ & 24.833 & 22.592 & 20.694 \\[1mm]
$35|1246$ & 28.794 & 25.927 & 23.390 \\[1mm]
$36|1245$ & 50.193 & 48.021 & 44.561 \\[1mm]
$45|1236$ & 30.629 & 28.950 & 27.092 \\[1mm]
$46|1235$ & 51.500 & 50.153 & 47.510 \\[1mm]
$56|1234$ & 56.080 & 55.390 & 51.521 \\[1mm]
$123|456$ & 56.661 & 55.474 & 54.345 \\[1mm]
$124|356$ & 54.402 & 52.666 & 52.044 \\[1mm]
$125|346$ & 50.653 & 48.953 & 47.993 \\[1mm]
$126|345$ & 28.957 & 26.470 & 22.905 \\[1mm]
$134|256$ & 47.675 & 46.956 & 46.154 \\[1mm]
$135|246$ & 47.279 & 43.109 & 40.016 \\[1mm]
$136|245$ & 34.237 & 29.400 & 26.527 \\[1mm]
$145|236$ & 47.331 & 42.957 & 41.199 \\[1mm]
$146|235$ & 35.340 & 34.072 & 30.970 \\[1mm]
$156|234$ & 39.487 & 38.426 & 36.880 \\
\bottomrule[0.6pt]
\end{tabular}
\caption{The comparison of the violation of the separability condition 
for the six-partite state considered in 
Ref.~\cite{PhysRevLett.114.050501}. Provided that the separability condition 
can be so strongly violated for all bipartitions it is absolutely unnecessary 
to test other kinds of separability, i.e. kinds with partitions of modes into 
three or more groups.}\label{tbl:6}
\end{table}

The last state considered in Ref.~\cite{PhysRevLett.114.050501} is a 
ten-partite state. It has been reported that the smallest violation of 1.1 was 
obtained for the bipartition $1,10|23456789$. The corresponding probability to 
get wrong result is $\mathbf{P}(1.1) \approx 0.27$, and it is not small enough 
to conclude that the state under study is not $1,10|23456789$-separable. 
Randomly generating vectors $\vec{h}$ and $\vec{g}$, we have found that the 
inequality \eqref{eq:E} for this kind of separability can be violated with 
$s = 3.65$. The corresponding probability $\mathbf{P}(3.65) < 3 \cdot 10^{-4}$ 
is much smaller and gives a strong confidence that the state is 
$1,10|23456789$-entangled. The vectors are
\begin{equation}
\begin{split}
    \vec{h} = (&-0.65, -1.22, -0.21, 0.01, 0.365, \\
            &-0.32, 0.25, 0.28, -0.88, -0.94), \\
    \vec{g} = (&0.22, -1.04, -0.12, -0.04, 0.42, \\
               &-0.34, 0.22, 0.30, -0.56, 0.73).
\end{split}
\end{equation}
The violations of other kinds of biseparability are all larger than 3, so the 
standard three-sigma test is passed for all bipartitions. The violations 
reported in Ref.~\cite{PhysRevLett.114.050501} show some strange behavior --- 
the violation for full separability is smaller than violations of some more 
coarse kinds. But this may be an artifact of an implementation of the genetic 
optimization algorithm.

Up to now it has been shown that the four-partite state with the covariance 
matrix \eqref{eq:cov4} is not separable for any fixed kind of separability. We 
demonstrate that this state is genuine entangled. To do this we need to 
find a pair of matrices $X$ and $P$ that simultaneously violate the 
inequalities 
\eqref{eq:E} for all bipartitions. First, we have tried to violate these 
inequalities with rank-one matrices. It took nearly one day, but we were able 
to find a pair of vectors
\begin{equation}
\begin{split}
    \vec{h} &= (0.31, -1.93, -0.17, -0.18) \\
    \vec{g} &= (0.30, 1.48, -0.57, -0.53),
\end{split}
\end{equation}
that violate the conditions \eqref{eq:E} for all seven bipartitions, and the 
minimal violation is $3.15$ (for the bipartition $1|234$). The corresponding 
probability $\mathbf{P}(3.15) = 1.6 \cdot 10^{-3}$ is relatively small to 
conclude that the state under study is genuine entangled.

The approach with a simpler conditions works but it takes a lot of time and it 
just marginally passes the three-sigma test. Using the general matrices we can 
do better. The sketch of our approach is as follows. We use a variant of the 
steepest gradient method.  According to this method, to optimize a convex 
function one has to go in the direction opposite to the gradient of the 
function. Here we have several functions to be optimized at once, and each has 
its own gradient. We start by generating a pair of random positive definite 
matrices $X$ and $P$ and compute the gradients of all seven target functions. 
If the directions of these gradients are not strongly scattered then we can 
take the average of the gradients, go in the opposite direction and still 
improving all our functions simultaneously. If the gradients point into nearly 
opposite directions then we cannot proceed this way, so we stop and generate a 
new random pair of matrices. We do this until we find proper matrices $X$ and 
$P$ or give up after some prescribed number of attempts. Following this 
approach, in a few hours we found the following pair of matrices for the 
four-partite state with the covariance matrix 
\eqref{eq:cov4}:
\begin{equation}
\begin{split}
    X &= 
    \begin{pmatrix}
        0.39234 & -0.20267 & 0.24691 & 0.30527 \\[1mm]
        -0.20267 & 0.88526 & 0.09450 & 0.09080 \\[1mm]
        0.24691 & 0.09450 & 0.58391 & 0.20795 \\[1mm]
        0.30527 & 0.09080 & 0.20795 & 0.39504
    \end{pmatrix} \\
    P &= 
    \begin{pmatrix}
        0.22992 & -0.13140 & -0.00477 & -0.11723 \\[1mm]
       -0.13140 & 0.52598 & -0.32316 & -0.16699 \\[1mm]
       -0.00477 & -0.32316 & 0.39949 & 0.06971 \\[1mm]
       -0.11723 & -0.16699 & 0.06971 & 0.31242
    \end{pmatrix} \nonumber
\end{split}
\end{equation}
For these matrices we have $\mathcal{G} = 1.47484$ and $\sigma = 
0.01947$. The bound $\mathcal{B}_{\mathcal{I}}(X, P)$ for 
different bipartitions is presented below. The elements of the matrices that 
were optimized over are highlighted. For the bipartition $1|234$ the maximum is 
attained at
\begin{equation}
\begin{split}
    X' &= 
    \begin{pmatrix}
        0.39234 & \tikzmarkin[hor=style orange]{el1}-0.10873 & 0.158136 & 
0.116524 \tikzmarkend{el1} \\[1mm]
        \tikzmarkin[ver=style orange]{el2}-0.10873 & 
0.88526 & 0.09450 & 0.09080 \\[1mm]
        0.158136 & 0.09450 & 0.58391 & 0.20795 \\[1mm]
        0.116524\tikzmarkend{el2} & 0.09080 & 0.20795 & 
0.39504
    \end{pmatrix} \\
    P' &= 
    \begin{pmatrix}
        0.22992 & \tikzmarkin[hor=style orange]{el3} -0.11914 & 0.113758 & 
0.083761\tikzmarkend{el3} \\[1mm]
       \tikzmarkin[hor=style orange]{el4}-0.11914 & 0.52598 & -0.32316 & 
-0.16699 \\[1mm]
       0.113758 & -0.32316 & 0.39949 & 0.06971 \\[1mm]
       0.083761\tikzmarkend{el4} & -0.16699 & 0.06971 & 0.31242
    \end{pmatrix}, \nonumber
\end{split}
\end{equation}
and is equal to $\mathcal{B}_{1|234}(X, P) = 1.65474$. For the bipartition 
$2|134$ at
\begin{equation}
\begin{split}
    X' &= 
    \begin{pmatrix}
        0.39234 & \tikzmarkin[hor=style orange]{el5}-0.07310\tikzmarkend{el5} 
& 0.24691 & 0.30527 \\[1mm]
        \tikzmarkin[hor=style orange]{el6}-0.07310\tikzmarkend{el6} & 
0.88526 & \tikzmarkin[hor=style orange]{el7}-0.03586 & 
-0.01993\tikzmarkend{el7} \\[1mm]
        0.24691 & \tikzmarkin[ver=style orange]{el8}-0.03586 & 0.58391 & 
0.20795 \\[1mm]
        0.30527 & -0.01993\tikzmarkend{el8} & 0.20795 & 0.39504
    \end{pmatrix} \\
    P' &= 
    \begin{pmatrix}
        0.22992 & \tikzmarkin[hor=style orange]{el9}-0.05400\tikzmarkend{el9} 
& -0.00477 & -0.11723 \\[1mm]
       \tikzmarkin[hor=style orange]{el10}-0.05400\tikzmarkend{el10} & 0.52598 
& \tikzmarkin[hor=style orange]{el11}-0.01432 & 0.02340\tikzmarkend{el11} 
\\[1mm]
       -0.00477 & \tikzmarkin[ver=style orange]{el12}-0.01432 & 0.39949 & 
0.06971 \\[1mm]
       -0.11723 & 0.02340\tikzmarkend{el12} & 0.06971 & 0.31242
    \end{pmatrix},\nonumber
\end{split}
\end{equation}
and is equal to $\mathcal{B}_{2|134}(X, P) = 1.66193$. For the bipartition 
$3|124$ at
\begin{equation}
\begin{split}
    X' &= 
    \begin{pmatrix}
        0.39234 & -0.20267 & \tikzmarkin[ver=style orange]{el13}0.22149 & 
0.30527 \\[1mm]
        -0.20267 & 0.88526 
& -0.01671\tikzmarkend{el13} & 0.09080  \\[1mm]
        \tikzmarkin[hor=style orange]{el14}0.22149 & -0.01671\tikzmarkend{el14} 
& 0.58391 & \tikzmarkin[hor=style orange]{el15}0.24154\tikzmarkend{el15} \\[1mm]
        0.30527 & 0.09080 & \tikzmarkin[hor=style 
orange]{el16}0.24154\tikzmarkend{el16} & 0.39504
    \end{pmatrix} \\
    P' &= 
    \begin{pmatrix}
        0.22992 & -0.13140 & \tikzmarkin[ver=style 
orange]{el17}0.02836 & -0.11723 \\[1mm]
       -0.13140 & 0.52598 & -0.07629\tikzmarkend{el17} & -0.16699 \\[1mm]
       \tikzmarkin[hor=style orange]{el18}0.02836 & -0.07629\tikzmarkend{el18} 
& 0.39949 & \tikzmarkin[hor=style orange]{el19}0.12842\tikzmarkend{el19} \\[1mm]
       -0.11723 & -0.16699 & \tikzmarkin[hor=style 
orange]{el20}0.12842\tikzmarkend{el20} & 0.31242
    \end{pmatrix}, \nonumber
\end{split}
\end{equation}
and is equal to $\mathcal{B}_{3|124}(X, P) = 1.56935$. For the bipartition 
$4|123$ at
\begin{equation}
\begin{split}
    X' &= 
    \begin{pmatrix}
        0.39234 & -0.20267 & 0.24691 & \tikzmarkin[ver=style 
orange]{el22}0.15483 \\[1mm]
        -0.20267 & 0.88526 & 0.09450 & 0.04047 \\[1mm]
        0.24691 & 0.09450 & 0.58391 & 0.23966\tikzmarkend{el22} \\[1mm]
        \tikzmarkin[hor=style orange]{el21}0.15483 & 0.04047 & 
0.23966\tikzmarkend{el21} & 0.39504
    \end{pmatrix} \\
    P' &= 
    \begin{pmatrix}
        0.22992 & -0.13140 & -0.00477 & \tikzmarkin[ver=style 
orange]{el24}0.05094 \\[1mm]
       -0.13140 & 0.52598 & -0.32316 & -0.05608 \\[1mm]
       -0.00477 & -0.32316 & 0.39949 & 0.12953\tikzmarkend{el24} \\[1mm]
       \tikzmarkin[hor=style orange]{el23}0.05094 & -0.05608
& 0.12953\tikzmarkend{el23}  & 0.31242
    \end{pmatrix}, \nonumber
\end{split}
\end{equation}
and is equal to $\mathcal{B}_{4|123}(X, P) = 1.63974$. For the bipartition 
$12|34$ at
\begin{equation}
\begin{split}
    X' &= 
    \begin{pmatrix}
        0.39234 & -0.20267 & \tikzmarkin[ver=style 
orange]{el25}0.19766 & 0.11649 \\[1mm]
        -0.20267 & 0.88526 & -0.02260 & 0.03156\tikzmarkend{el25} \\[1mm]
        \tikzmarkin[ver=style orange]{el26}0.19766 & -0.02260 & 0.58391 & 
0.20795 \\[1mm]
        0.11649 & 0.03156\tikzmarkend{el26} & 0.20795 & 0.39504
    \end{pmatrix} \\
    P' &= 
    \begin{pmatrix}
        0.22992 & -0.13140 & \tikzmarkin[ver=style 
orange]{el27}0.11949 & 0.06522 \\[1mm]
       -0.13140 & 0.52598 & -0.02001 & 0.02362\tikzmarkend{el27} \\[1mm]
       \tikzmarkin[ver=style 
orange]{el28}0.11949 & -0.02001 & 0.39949 & 0.06971 \\[1mm]
       0.06522 & 0.02362\tikzmarkend{el28} & 0.06971 & 0.31242
    \end{pmatrix}, \nonumber
\end{split}
\end{equation}
and is equal to $\mathcal{B}_{12|34}(X, P) = 1.81056$. For the bipartition 
$13|24$ at
\begin{equation}
\begin{split}
    X' &= 
    \begin{pmatrix}
        0.39234 & \tikzmarkin[hor=style 
orange]{el29}-0.10013\tikzmarkend{el29} & 0.24691 & \tikzmarkin[hor=style 
orange]{el30}0.12997\tikzmarkend{el30} \\[1mm]
        \tikzmarkin[hor=style orange]{el31}-0.10013\tikzmarkend{el31} & 0.88526 
& \tikzmarkin[hor=style orange]{el32}-0.03695\tikzmarkend{el32} & 
0.09080 \\[1mm]
        0.24691 & \tikzmarkin[hor=style orange]{el33}-0.03695\tikzmarkend{el33} 
& 0.58391 & \tikzmarkin[hor=style orange]{el34}0.25436\tikzmarkend{el34} 
\\[1mm]
        \tikzmarkin[hor=style orange]{el35}0.12997\tikzmarkend{el35} & 0.09080 
& \tikzmarkin[hor=style orange]{el36}0.25436\tikzmarkend{el36} & 0.39504
    \end{pmatrix} \\
    P' &= 
    \begin{pmatrix}
        0.22992 & \tikzmarkin[hor=style 
orange]{el37}-0.07273\tikzmarkend{el37} & -0.00477 & \tikzmarkin[hor=style 
orange]{el38}0.06225\tikzmarkend{el38} \\[1mm]
       \tikzmarkin[hor=style orange]{el39}-0.07273\tikzmarkend{el39} & 0.52598 
& \tikzmarkin[hor=style orange]{el40}-0.05209\tikzmarkend{el40} & -0.16699 
\\[1mm]
       -0.00477 & \tikzmarkin[hor=style orange]{el41}-0.05209\tikzmarkend{el41} 
& 0.39949 & \tikzmarkin[hor=style orange]{el42}0.17734\tikzmarkend{el42} \\[1mm]
       \tikzmarkin[hor=style orange]{el43}0.06225\tikzmarkend{el43} & -0.16699 
& \tikzmarkin[hor=style orange]{el44}0.17734\tikzmarkend{el44} & 0.31242
    \end{pmatrix}, \nonumber
\end{split}
\end{equation}
and is equal to $\mathcal{B}_{13|24}(X, P) = 1.74993$. For the bipartition 
$14|23$ at
\begin{equation}
\begin{split}
    X' &= 
    \begin{pmatrix}
        0.39234 & \tikzmarkin[hor=style orange]{el45}-0.11435 & 
0.18571\tikzmarkend{el45} & 0.30527 \\[1mm]
        \tikzmarkin[ver=style orange]{el47}-0.11435 & 0.88526 & 0.09450 & 
\tikzmarkin[ver=style orange]{el48}-0.02360 \\[1mm]
        0.18571\tikzmarkend{el47} & 0.09450 & 0.58391 & 
0.25307\tikzmarkend{el48} \\[1mm]
        0.30527 & \tikzmarkin[hor=style orange]{el46}-0.02360 & 
0.25307\tikzmarkend{el46} & 0.39504
    \end{pmatrix} \\
    P' &= 
    \begin{pmatrix}
        0.22992 & \tikzmarkin[hor=style orange]{el49}-0.09531 & 
0.05497\tikzmarkend{el49} & -0.11723 \\[1mm]
       \tikzmarkin[ver=style orange]{el51}-0.09531 & 0.52598 & -0.32316 & 
\tikzmarkin[ver=style orange]{el52}-0.02171 \\[1mm]
       0.05497\tikzmarkend{el51} & -0.32316 & 0.39949 & 
0.12643\tikzmarkend{el52} \\[1mm]
       -0.11723 & \tikzmarkin[hor=style orange]{el50}-0.02171 & 
0.12643\tikzmarkend{el50} & 0.31242
    \end{pmatrix}, \nonumber
\end{split}
\end{equation}
and is equal to $\mathcal{B}_{14|23}(X, P) = 1.56114$. The smallest number among 
these maximums is the last one, $1.56114$, so we have
\begin{equation}
    \frac{\mathcal{B}_{\mathcal{I}}(X, P) - \mathcal{G}(X, P)}{\sigma(X, P)} 
    \geq s_0 = 4.43199 \nonumber
\end{equation}
for all bipartitions $\mathcal{I}$ simultaneously. The corresponding 
probability is $\mathbf{P}(s_0) < 10^{-5}$, which is almost two orders of 
magnitude smaller than for the vectors $\vec{h}$ and $\vec{g}$ we found before, 
so one can be pretty sure that the state under study is genuine entangled.


\begin{thebibliography}{23}%
\makeatletter
\providecommand \@ifxundefined [1]{%
 \@ifx{#1\undefined}
}%
\providecommand \@ifnum [1]{%
 \ifnum #1\expandafter \@firstoftwo
 \else \expandafter \@secondoftwo
 \fi
}%
\providecommand \@ifx [1]{%
 \ifx #1\expandafter \@firstoftwo
 \else \expandafter \@secondoftwo
 \fi
}%
\providecommand \natexlab [1]{#1}%
\providecommand \enquote  [1]{``#1''}%
\providecommand \bibnamefont  [1]{#1}%
\providecommand \bibfnamefont [1]{#1}%
\providecommand \citenamefont [1]{#1}%
\providecommand \href@noop [0]{\@secondoftwo}%
\providecommand \href [0]{\begingroup \@sanitize@url \@href}%
\providecommand \@href[1]{\@@startlink{#1}\@@href}%
\providecommand \@@href[1]{\endgroup#1\@@endlink}%
\providecommand \@sanitize@url [0]{\catcode `\\12\catcode `\$12\catcode
  `\&12\catcode `\#12\catcode `\^12\catcode `\_12\catcode `\%12\relax}%
\providecommand \@@startlink[1]{}%
\providecommand \@@endlink[0]{}%
\providecommand \url  [0]{\begingroup\@sanitize@url \@url }%
\providecommand \@url [1]{\endgroup\@href {#1}{\urlprefix }}%
\providecommand \urlprefix  [0]{URL }%
\providecommand \Eprint [0]{\href }%
\providecommand \doibase [0]{http://dx.doi.org/}%
\providecommand \selectlanguage [0]{\@gobble}%
\providecommand \bibinfo  [0]{\@secondoftwo}%
\providecommand \bibfield  [0]{\@secondoftwo}%
\providecommand \translation [1]{[#1]}%
\providecommand \BibitemOpen [0]{}%
\providecommand \bibitemStop [0]{}%
\providecommand \bibitemNoStop [0]{.\EOS\space}%
\providecommand \EOS [0]{\spacefactor3000\relax}%
\providecommand \BibitemShut  [1]{\csname bibitem#1\endcsname}%
\let\auto@bib@innerbib\@empty
\bibitem [{\citenamefont {Braunstein}\ and\ \citenamefont {van
  Loock}(2005)}]{RevModPhys.77.513}%
  \BibitemOpen
  \bibfield  {author} {\bibinfo {author} {\bibfnamefont {S.~L.}\ \bibnamefont
  {Braunstein}}\ and\ \bibinfo {author} {\bibfnamefont {P.}~\bibnamefont {van
  Loock}},\ }\href@noop {} {\bibfield  {journal} {\bibinfo  {journal} {Rev.
  Mod. Phys.}\ }\textbf {\bibinfo {volume} {77}},\ \bibinfo {pages} {513}
  (\bibinfo {year} {2005})}\BibitemShut {NoStop}%
\bibitem [{\citenamefont {Weedbrook}\ \emph {et~al.}(2012)\citenamefont
  {Weedbrook}, \citenamefont {Pirandola}, \citenamefont {Garc\'{i}a-Patr\'on},
  \citenamefont {Cerf}, \citenamefont {Ralph}, \citenamefont {Shapiro},\ and\
  \citenamefont {Lloyd}}]{RevModPhys.84.621}%
  \BibitemOpen
  \bibfield  {author} {\bibinfo {author} {\bibfnamefont {C.}~\bibnamefont
  {Weedbrook}}, \bibinfo {author} {\bibfnamefont {S.}~\bibnamefont
  {Pirandola}}, \bibinfo {author} {\bibfnamefont {R.}~\bibnamefont
  {Garc\'{i}a-Patr\'on}}, \bibinfo {author} {\bibfnamefont {N.~J.}\
  \bibnamefont {Cerf}}, \bibinfo {author} {\bibfnamefont {T.~C.}\ \bibnamefont
  {Ralph}}, \bibinfo {author} {\bibfnamefont {J.~H.}\ \bibnamefont {Shapiro}},
  \ and\ \bibinfo {author} {\bibfnamefont {S.}~\bibnamefont {Lloyd}},\
  }\href@noop {} {\bibfield  {journal} {\bibinfo  {journal} {Rev. Mod. Phys.}\
  }\textbf {\bibinfo {volume} {84}},\ \bibinfo {pages} {621} (\bibinfo {year}
  {2012})}\BibitemShut {NoStop}%
\bibitem [{\citenamefont {Duan}\ \emph {et~al.}(2000)\citenamefont {Duan},
  \citenamefont {Giedke}, \citenamefont {Cirac},\ and\ \citenamefont
  {Zoller}}]{PhysRevLett.84.2722}%
  \BibitemOpen
  \bibfield  {author} {\bibinfo {author} {\bibfnamefont {L.-M.}\ \bibnamefont
  {Duan}}, \bibinfo {author} {\bibfnamefont {G.}~\bibnamefont {Giedke}},
  \bibinfo {author} {\bibfnamefont {J.~I.}\ \bibnamefont {Cirac}}, \ and\
  \bibinfo {author} {\bibfnamefont {P.}~\bibnamefont {Zoller}},\ }\href@noop {}
  {\bibfield  {journal} {\bibinfo  {journal} {Phys. Rev. Lett.}\ }\textbf
  {\bibinfo {volume} {84}},\ \bibinfo {pages} {2722} (\bibinfo {year}
  {2000})}\BibitemShut {NoStop}%
\bibitem [{\citenamefont {van Loock}\ and\ \citenamefont
  {Furusawa}(2003)}]{PhysRevA.67.052315}%
  \BibitemOpen
  \bibfield  {author} {\bibinfo {author} {\bibfnamefont {P.}~\bibnamefont {van
  Loock}}\ and\ \bibinfo {author} {\bibfnamefont {A.}~\bibnamefont
  {Furusawa}},\ }\href@noop {} {\bibfield  {journal} {\bibinfo  {journal}
  {Phys. Rev. A}\ }\textbf {\bibinfo {volume} {67}},\ \bibinfo {pages} {052315}
  (\bibinfo {year} {2003})}\BibitemShut {NoStop}%
\bibitem [{\citenamefont {Teh}\ and\ \citenamefont
  {Reid}(2014)}]{PhysRevA.90.062337}%
  \BibitemOpen
  \bibfield  {author} {\bibinfo {author} {\bibfnamefont {R.~Y.}\ \bibnamefont
  {Teh}}\ and\ \bibinfo {author} {\bibfnamefont {M.~D.}\ \bibnamefont {Reid}},\
  }\href@noop {} {\bibfield  {journal} {\bibinfo  {journal} {Phys. Rev. A}\
  }\textbf {\bibinfo {volume} {90}},\ \bibinfo {pages} {062337} (\bibinfo
  {year} {2014})}\BibitemShut {NoStop}%
\bibitem [{\citenamefont {Shchukin}\ and\ \citenamefont {van
  Loock}(2014)}]{PhysRevA.90.012334}%
  \BibitemOpen
  \bibfield  {author} {\bibinfo {author} {\bibfnamefont {E.}~\bibnamefont
  {Shchukin}}\ and\ \bibinfo {author} {\bibfnamefont {P.}~\bibnamefont {van
  Loock}},\ }\href@noop {} {\bibfield  {journal} {\bibinfo  {journal} {Phys.
  Rev. A}\ }\textbf {\bibinfo {volume} {90}},\ \bibinfo {pages} {012334}
  (\bibinfo {year} {2014})}\BibitemShut {NoStop}%
\bibitem [{\citenamefont {Sperling}\ and\ \citenamefont
  {Vogel}(2013)}]{PhysRevLett.111.110503}%
  \BibitemOpen
  \bibfield  {author} {\bibinfo {author} {\bibfnamefont {J.}~\bibnamefont
  {Sperling}}\ and\ \bibinfo {author} {\bibfnamefont {W.}~\bibnamefont
  {Vogel}},\ }\href@noop {} {\bibfield  {journal} {\bibinfo  {journal} {Phys.
  Rev. Lett.}\ }\textbf {\bibinfo {volume} {111}},\ \bibinfo {pages} {110503}
  (\bibinfo {year} {2013})}\BibitemShut {NoStop}%
\bibitem [{\citenamefont {Gerke}\ \emph {et~al.}(2015)\citenamefont {Gerke},
  \citenamefont {Sperling}, \citenamefont {Vogel}, \citenamefont {Cai},
  \citenamefont {Roslund}, \citenamefont {Treps},\ and\ \citenamefont
  {Fabre}}]{PhysRevLett.114.050501}%
  \BibitemOpen
  \bibfield  {author} {\bibinfo {author} {\bibfnamefont {S.}~\bibnamefont
  {Gerke}}, \bibinfo {author} {\bibfnamefont {J.}~\bibnamefont {Sperling}},
  \bibinfo {author} {\bibfnamefont {W.}~\bibnamefont {Vogel}}, \bibinfo
  {author} {\bibfnamefont {Y.}~\bibnamefont {Cai}}, \bibinfo {author}
  {\bibfnamefont {J.}~\bibnamefont {Roslund}}, \bibinfo {author} {\bibfnamefont
  {N.}~\bibnamefont {Treps}}, \ and\ \bibinfo {author} {\bibfnamefont
  {C.}~\bibnamefont {Fabre}},\ }\href@noop {} {\bibfield  {journal} {\bibinfo
  {journal} {Phys. Rev. Lett.}\ }\textbf {\bibinfo {volume} {114}},\ \bibinfo
  {pages} {050501} (\bibinfo {year} {2015})}\BibitemShut {NoStop}%
\bibitem [{\citenamefont {Hyllus}\ and\ \citenamefont
  {Eisert}(2006)}]{NewJPhys.8.51}%
  \BibitemOpen
  \bibfield  {author} {\bibinfo {author} {\bibfnamefont {P.}~\bibnamefont
  {Hyllus}}\ and\ \bibinfo {author} {\bibfnamefont {J.}~\bibnamefont
  {Eisert}},\ }\href@noop {} {\bibfield  {journal} {\bibinfo  {journal} {New
  Journal of Physics}\ }\textbf {\bibinfo {volume} {8}},\ \bibinfo {pages} {51}
  (\bibinfo {year} {2006})}\BibitemShut {NoStop}%
\bibitem [{\citenamefont {Shalm}\ \emph {et~al.}(2013)\citenamefont {Shalm},
  \citenamefont {Hamel}, \citenamefont {Yan}, \citenamefont {Simon},
  \citenamefont {Resch},\ and\ \citenamefont {Jennewein}}]{NatPhys.9.19}%
  \BibitemOpen
  \bibfield  {author} {\bibinfo {author} {\bibfnamefont {L.~K.}\ \bibnamefont
  {Shalm}}, \bibinfo {author} {\bibfnamefont {D.~R.}\ \bibnamefont {Hamel}},
  \bibinfo {author} {\bibfnamefont {Z.}~\bibnamefont {Yan}}, \bibinfo {author}
  {\bibfnamefont {C.}~\bibnamefont {Simon}}, \bibinfo {author} {\bibfnamefont
  {K.~J.}\ \bibnamefont {Resch}}, \ and\ \bibinfo {author} {\bibfnamefont
  {T.}~\bibnamefont {Jennewein}},\ }\href@noop {} {\bibfield  {journal}
  {\bibinfo  {journal} {Nat. Phys.}\ }\textbf {\bibinfo {volume} {9}},\
  \bibinfo {pages} {19} (\bibinfo {year} {2013})}\BibitemShut {NoStop}%
\bibitem [{\citenamefont {Valido}\ \emph {et~al.}(2014)\citenamefont {Valido},
  \citenamefont {Levi},\ and\ \citenamefont {Mintert}}]{PhysRevA.90.052321}%
  \BibitemOpen
  \bibfield  {author} {\bibinfo {author} {\bibfnamefont {A.~A.}\ \bibnamefont
  {Valido}}, \bibinfo {author} {\bibfnamefont {F.}~\bibnamefont {Levi}}, \ and\
  \bibinfo {author} {\bibfnamefont {F.}~\bibnamefont {Mintert}},\ }\href@noop
  {} {\bibfield  {journal} {\bibinfo  {journal} {Phys. Rev. A}\ }\textbf
  {\bibinfo {volume} {90}},\ \bibinfo {pages} {052321} (\bibinfo {year}
  {2014})}\BibitemShut {NoStop}%
\bibitem [{\citenamefont {Eisert}\ \emph {et~al.}(2004)\citenamefont {Eisert},
  \citenamefont {Hyllus}, \citenamefont {G\"uhne},\ and\ \citenamefont
  {Curty}}]{PhysRevA.70.062317}%
  \BibitemOpen
  \bibfield  {author} {\bibinfo {author} {\bibfnamefont {J.}~\bibnamefont
  {Eisert}}, \bibinfo {author} {\bibfnamefont {P.}~\bibnamefont {Hyllus}},
  \bibinfo {author} {\bibfnamefont {O.}~\bibnamefont {G\"uhne}}, \ and\
  \bibinfo {author} {\bibfnamefont {M.}~\bibnamefont {Curty}},\ }\href@noop {}
  {\bibfield  {journal} {\bibinfo  {journal} {Phys. Rev. A}\ }\textbf {\bibinfo
  {volume} {70}},\ \bibinfo {pages} {062317} (\bibinfo {year}
  {2004})}\BibitemShut {NoStop}%
\bibitem [{\citenamefont {de~Gosson}(2006)}]{SGQM}%
  \BibitemOpen
  \bibfield  {author} {\bibinfo {author} {\bibfnamefont {M.~A.}\ \bibnamefont
  {de~Gosson}},\ }\href@noop {} {\emph {\bibinfo {title} {Symplectic Geometry
  and Quantum Mechanics}}}\ (\bibinfo  {publisher} {Birkh\"{a}user},\ \bibinfo
  {year} {2006})\BibitemShut {NoStop}%
\bibitem [{\citenamefont {Simon}\ \emph {et~al.}(1988)\citenamefont {Simon},
  \citenamefont {Sudarshan},\ and\ \citenamefont {Mukunda}}]{PhysRevA.37.3028}%
  \BibitemOpen
  \bibfield  {author} {\bibinfo {author} {\bibfnamefont {R.}~\bibnamefont
  {Simon}}, \bibinfo {author} {\bibfnamefont {E.~C.~G.}\ \bibnamefont
  {Sudarshan}}, \ and\ \bibinfo {author} {\bibfnamefont {N.}~\bibnamefont
  {Mukunda}},\ }\href@noop {} {\bibfield  {journal} {\bibinfo  {journal} {Phys.
  Rev. A}\ }\textbf {\bibinfo {volume} {37}},\ \bibinfo {pages} {3028}
  (\bibinfo {year} {1988})}\BibitemShut {NoStop}%
\bibitem [{\citenamefont {Horn}\ and\ \citenamefont
  {Johnson}(2013)}]{horn-johnson}%
  \BibitemOpen
  \bibfield  {author} {\bibinfo {author} {\bibfnamefont {R.~A.}\ \bibnamefont
  {Horn}}\ and\ \bibinfo {author} {\bibfnamefont {C.~R.}\ \bibnamefont
  {Johnson}},\ }\href@noop {} {\emph {\bibinfo {title} {Matrix Analysis}}},\
  \bibinfo {edition} {2nd}\ ed.\ (\bibinfo  {publisher} {Cambridge University
  Press},\ \bibinfo {year} {2013})\BibitemShut {NoStop}%
\bibitem [{\citenamefont {Lieb}\ and\ \citenamefont
  {Thirring}(1975)}]{PhysRevLett.35.687}%
  \BibitemOpen
  \bibfield  {author} {\bibinfo {author} {\bibfnamefont {E.~H.}\ \bibnamefont
  {Lieb}}\ and\ \bibinfo {author} {\bibfnamefont {W.~E.}\ \bibnamefont
  {Thirring}},\ }\href@noop {} {\bibfield  {journal} {\bibinfo  {journal}
  {Phys. Rev. Lett.}\ }\textbf {\bibinfo {volume} {35}},\ \bibinfo {pages}
  {687} (\bibinfo {year} {1975})}\BibitemShut {NoStop}%
\bibitem [{\citenamefont {Lieb}\ and\ \citenamefont
  {Thirring}(1976)}]{lieb-thirring}%
  \BibitemOpen
  \bibfield  {author} {\bibinfo {author} {\bibfnamefont {E.~H.}\ \bibnamefont
  {Lieb}}\ and\ \bibinfo {author} {\bibfnamefont {W.~E.}\ \bibnamefont
  {Thirring}},\ }in\ \href@noop {} {\emph {\bibinfo {booktitle} {Studies in
  Mathematical Physics}}},\ \bibinfo {editor} {edited by\ \bibinfo {editor}
  {\bibfnamefont {E.}~\bibnamefont {Lieb}}, \bibinfo {editor} {\bibfnamefont
  {B.}~\bibnamefont {Simon}}, \ and\ \bibinfo {editor} {\bibfnamefont
  {A.}~\bibnamefont {Wightman}}}\ (\bibinfo  {publisher} {Princeton University
  Press},\ \bibinfo {year} {1976})\ pp.\ \bibinfo {pages}
  {269--303}\BibitemShut {NoStop}%
\bibitem [{\citenamefont {Araki}(1990)}]{Lett.Math.Phys.19.167}%
  \BibitemOpen
  \bibfield  {author} {\bibinfo {author} {\bibfnamefont {H.}~\bibnamefont
  {Araki}},\ }\href@noop {} {\bibfield  {journal} {\bibinfo  {journal} {Lett.
  Math. Phys.}\ }\textbf {\bibinfo {volume} {19}},\ \bibinfo {pages} {167}
  (\bibinfo {year} {1990})}\BibitemShut {NoStop}%
\bibitem [{\citenamefont {Ohya}\ and\ \citenamefont {Petz}(1993)}]{entropy}%
  \BibitemOpen
  \bibfield  {author} {\bibinfo {author} {\bibfnamefont {M.}~\bibnamefont
  {Ohya}}\ and\ \bibinfo {author} {\bibfnamefont {D.}~\bibnamefont {Petz}},\
  }\href@noop {} {\emph {\bibinfo {title} {Quantum entropy and its use}}}\
  (\bibinfo  {publisher} {Springer-Verlag},\ \bibinfo {year}
  {1993})\BibitemShut {NoStop}%
\bibitem [{\citenamefont {Werner}\ and\ \citenamefont
  {Wolf}(2001)}]{PhysRevLett.86.3658}%
  \BibitemOpen
  \bibfield  {author} {\bibinfo {author} {\bibfnamefont {R.~F.}\ \bibnamefont
  {Werner}}\ and\ \bibinfo {author} {\bibfnamefont {M.~M.}\ \bibnamefont
  {Wolf}},\ }\href@noop {} {\bibfield  {journal} {\bibinfo  {journal} {Phys.
  Rev. Lett.}\ }\textbf {\bibinfo {volume} {86}},\ \bibinfo {pages} {3658}
  (\bibinfo {year} {2001})}\BibitemShut {NoStop}%
\bibitem [{Note1()}]{Note1}%
  \BibitemOpen
  \bibinfo {note} {In fact, in Ref.~\cite {PhysRevA.67.052315} the term
  \protect \textit {genuine entanglement} was used in a different, weaker,
  sense. But the condition obtained there happens to work for genuine
  entanglement in the present, stronger, sense.}\BibitemShut {Stop}%
\bibitem [{\citenamefont {Boyd}\ and\ \citenamefont
  {Vandenberghe}(2004)}]{convex-opt}%
  \BibitemOpen
  \bibfield  {author} {\bibinfo {author} {\bibfnamefont {S.}~\bibnamefont
  {Boyd}}\ and\ \bibinfo {author} {\bibfnamefont {L.}~\bibnamefont
  {Vandenberghe}},\ }\href@noop {} {\emph {\bibinfo {title} {Convex
  Optimization}}}\ (\bibinfo  {publisher} {Cambridge University Press},\
  \bibinfo {year} {2004})\BibitemShut {NoStop}%
\bibitem [{\citenamefont {Grafarend}(2006)}]{lnm}%
  \BibitemOpen
  \bibfield  {author} {\bibinfo {author} {\bibfnamefont {E.~W.}\ \bibnamefont
  {Grafarend}},\ }\href@noop {} {\emph {\bibinfo {title} {Linear and Nonlinear
  Models: Fixed Effects, Random Effects, and Mixed Models}}}\ (\bibinfo
  {publisher} {Walter de Gruyter},\ \bibinfo {year} {2006})\BibitemShut
  {NoStop}%
\end{thebibliography}
\end{document}